\documentclass[preprint,10pt,authoryear]{elsarticle}
\usepackage{amssymb}
\usepackage{lineno}
\usepackage{natbib}
\usepackage{placeins}
\usepackage{amsmath, nccmath}
\usepackage{geometry}
\usepackage{mathtools}
\usepackage{ragged2e}
\makeatletter
\newcommand{\vast}{\bBigg@{3}}
\newcommand{\Vast}{\bBigg@{4}}
\makeatother
\usepackage{booktabs,xcolor,siunitx}
\definecolor{lightgray}{gray}{0.9}
\usepackage[bookmarks, colorlinks, breaklinks]{hyperref}  
\definecolor{midnightblue}{rgb}{0.0, 0.49, 0.67}
\hypersetup{linkcolor=midnightblue,citecolor=midnightblue,filecolor=black,urlcolor=midnightblue} 
\usepackage{etoolbox}
\usepackage[T1]{fontenc}
\usepackage{lmodern}

\journal{Preprint submitted to arXiv}
\begin{document}
\begin{frontmatter}
\title{\textbf{Hydro-mechanical earthquake cycles in a poro-visco-elasto-plastic fluid-bearing fault structure}}
\author{Luca Dal Zilio$^{ 1 *}$, Betti Hegyi$^{ 2 \,}$, Whitney Behr$^{ 2 \,}$, and Taras Gerya$^{ 3}$}
\address{$^1$ Seismology and Geodynamics, Institute of Geophysics, Department of Earth Sciences, Swiss Federal Institute of Technology (ETH Z\"urich), Z\"urich, Switzerland}
\address{$^2$ Structural Geology and Tectonics, Geological Institute, Department of Earth Sciences, Swiss Federal Institute of Technology (ETH Z\"urich), Z\"urich, Switzerland}
\address{$^3$ Geophysical Fluid Dynamics, Institute of Geophysics, Department of Earth Sciences, Swiss Federal Institute of Technology (ETH Z\"urich), Z\"urich, Switzerland}
\address{* Corresponding authors: L. Dal Zilio, luca.dalzilio@erdw.ethz.ch \\[15pt]}

\address{\justifying{\par\noindent\rule{\textwidth}{0.4pt} \\[4pt] 

\noindent\normalfont\normalsize This manuscript is a \textbf{preprint} uploaded to \textbf{arXiv.org}. The first version of this preprint has been submitted for publication on the 27th of January 2022. Authors encourage downloading the latest manuscript version from arXiv.org, and welcome comments, feedback and discussions anytime. \\[-30pt]}}
\begin{abstract}
A major goal in earthquake physics is to derive a constitutive framework for fault slip that captures the dependence of shear strength on fault rheology, sliding velocity, and pore-fluid pressure. In this study, we present \texttt{H-MEC} (Hydro-Mechanical Earthquake Cycles), a newly-developed two-phase flow numerical code --- which couples solid rock deformation and pervasive fluid flow --- to simulate how crustal stress and fluid pressure evolve during the earthquake cycle on a fluid-bearing fault structure. This unified, continuum-based model, incorporates a staggered finite difference--marker-in-cell (SFD-MIC) method and accounts for full inertial (wave mediated) effects and fluid flow in poro-visco-elasto-plastic compressible medium. Global Picard-iterations and an adaptive time stepping allows the correct resolution of both long- and short-time scales, ranging from years during slow tectonic loading to milliseconds during the propagation of dynamic ruptures. We present a comprehensive in-plane strike-slip setup in which we test analytical poroelastic benchmarks of pore-fluid pressure diffusion from an injection point along a finite fault width. We then investigate how pore-fluid pressure evolution and solid--fluid compressibility control sequences of \textit{seismic} and \textit{aseismic} slip on geologic faults. While the onset of fluid-driven shear cracks is controlled by localized collapse of pores and dynamic self-pressurization of fluids inside the undrained fault zone, subsequent dynamic ruptures are driven by solitary pulse-like fluid pressure waves propagating at seismic speed. Furthermore, shear strength weakening associated with rapid self-pressurization of pore-fluid can account for the slip--fracture energy scaling observed for large earthquakes. This numerical framework provides a viable tool to better understand fluid-driven dynamic ruptures --- either as a natural process or induced by human activities --- and highlights the importance of considering the realistic hydro-mechanical structure of faults to investigate sequences of \textit{seismic} and \textit{aseismic} slip.
\end{abstract}
\begin{keyword}
\texttt{Rock-Fluid Interaction \sep Earthquake Cycles \sep Fault Mechanics \sep Poro-Visco-Elasto-Plastic Rheology \sep Finite Difference}
\end{keyword}
\end{frontmatter}
\setlength\parindent{0pt}
\section{Introduction}  
There is a growing interest in understanding how geologic faults respond to transient sources of fluid. This issue has become a societal concern due to the increasing interest in geothermal energy \citep[e.g.,][]{grigoli2017current,terakawa2012high}, CO$_2$ storage \citep{zoback2012earthquake}, or injection of waste waters associated with hydraulic fracking \citep{ellsworth2013injection}. Natural and artificial sources of fluid can elevate pore-fluid which can destabilize active faults through either slow-slip or fast, seismic-wave-producing rupture events perceived as earthquakes. Examples of fluid-driven seismic and aseismic slip on natural faults are related to seismic swarms \citep[e.g.,][]{roland2009earthquake,ross20203d,chen2012spatial,hatch2020evidence,hainzl2004seismicity}, aftershock sequences \citep[e.g.,][]{miller2020aftershocks,nur1972aftershocks,ross2017aftershocks,miller2004aftershocks}, and other natural phenomena such as slow-slip events and tectonic tremors \citep[e.g.,][]{jolivet2020transient,burgmann2018geophysics,behr2021s}. On the other hand, fluid injections associated with human activities are well known to lead to the reactivation of faults, slow-slip transients, and earthquakes, which occasionally are large enough to cause damage \citep[e.g.,][]{guglielmi2015seismicity,deichmann2009earthquakes,ellsworth2019triggering,keranen2013potentially,raleigh1976experiment,keranen2014sharp,bao2016fault}. Taken together, this body of work suggest that fluid-driven seismic and aseismic fault slip in nature and anthropogenic fluid injections are closely linked. \\

The propagation of fluid-driven shear cracks, the evolution of pore pressure, and the poroelastic response of faults, have been subjects of interest for decades in the earthquake modeling community \citep[e.g.,][]{rice1976some,rice1992fault,dunham2008earthquake,rudnicki2006effective,cruz2018rapid,heimisson2019poroelastic,zhu2020fault,petrini2020seismo}. Recent modeling efforts have mostly focused on the stability of frictional slip when a fluid is locally injected and diffuses within a fault \citep{bhattacharya2019fluid,cappa2018relationship,garagash2012nucleation,larochelle2021constraining,yang2021effect}, or when fluid injection initiates shear cracks on a strengthening rate-and-state frictional fault \citep{dublanchet2019fluid}. Despite the apparent relevance of fluid-driven slow and fast slip transients in a wide variety of natural and anthropogenic environments, the spatio-temporal evolution of sequences of \textit{seismic} and \textit{aseismic} slip in response to pore-fluid evolution remains poorly constrained. This is, in part, due to the challenge of solving a fully coupled solid-fluid system in which the dynamic evolution of fault slip, stress, pore-fluid pressure, and shear strength are unknown. \\

In this study, we present H-MEC (Hydro-Mechanical Earthquake Cycles), a continuum-based modeling approach to simulate fully dynamic sequences of seismic and aseismic slip (SEAS) in a poro-visco-elasto-plastic compressible medium with rate-dependent strength, which considerably extends our previous numerical model developed for incompressible media with rate-independent strength \citep{petrini2020seismo}. The computational method is developed for a plane-strain problem on a simplified strike-slip fault. Our coupled formulation employs a Picard-type non-linear solver with a fully implicit, first order accurate time integrator that utilizes an adaptive time stepping to efficiently solve the system through multiple seismic cycles. We apply a staggered finite difference--marker-in-cell (SFD-MIC) method, as it is easy to program, efficient, and the marker-in-cell technique can be applied to simulate highly deforming materials, such as faults and geologic structures \citep{gerya2007robust,petrini2020seismo}. \\

The purpose of this study is to introduce a quantitative set of numerical experiments in which we test analytical benchmarks and illustrative examples of sequences of \textit{seismic} and \textit{aseismic} slip along with results from the solid-fluid coupling, fluid transport on- and off-fault, pore pressure evolution, and fault slip. Our focus is to demonstrate the validity of our methodology and highlight the mechanical processes and phenomena that arise from this solid-fluid coupling in a generic sense. We apply this methodology in the context of a simple, 2-D plane-strain shear model of a strike-slip fault in a uniform bulk domain, which is an idealization of a natural strike-slip fault. While parameter choices are chosen to be representative of a strike-slip fault, we are not attempting to model any specific fault or earthquake sequence.  In Section \ref{methodology}, we present the conceptual and mathematical model and state the continuum problem solved in this work. In Section \ref{results}, we present results of two analytical benchmarks and a set of simulations of earthquake sequences. Finally, Section \ref{discussion} we analyse and interpret the results and compare solutions for different confining pressure, solid and fluid compressibility, porosity, and fracture energy. Lastly, we discuss future avenues for research through the use of this methodology.

\section{Methodology}\label{methodology}
In this section we present the conceptual, mathematical, and physical foundations of the model before presenting the numerical simulations that will follow.
\subsection{Governing equations}
 The mathematical description of a deforming continuum in a stationary gravity field is given by the following set of conservation equations for total momentum (solid matrix and fluid; Eq. \ref{mom-sol+flu}), fluid momentum (Eq. \ref{mom-flu}), fully compressible solid mass (Eq. \ref{mass-sol}), and fully compressible fluid mass (Eq. \ref{mass-flu}):

\begin{fleqn}[\parindent]
\begin{align} 
\nabla \cdot \boldsymbol{\underline{\underline{\sigma}}} + \boldsymbol{g}  \, \rho^{[\text{t}]} = \, \rho^{[\text{t}]} \frac{D^{[\text{s}]} \boldsymbol{v}^{[\text{s}]}}{D t} \; , \label{mom-sol+flu}
\end{align} 
\end{fleqn}
\begin{fleqn}[\parindent]
\begin{align}
\boldsymbol{v}^{[\text{D}]} = - \frac{k^{[\phi]}}{\eta^{[\text{f\,}]}} \; \bigg(\nabla p^{[\text{f\,}]} - \rho^{[\text{f\,}]} \, \bigg( \boldsymbol{g} - \frac{D^{[\text{f\,}]} \boldsymbol{v}^{[\text{f\,}]}}{D t} \bigg) \bigg) \; ,
\label{mom-flu} 
\end{align}
\end{fleqn}
\begin{fleqn}[\parindent]
\begin{align}
\nabla \cdot \boldsymbol{v}^{[\text{s}]} = - \frac{1}{K^{[\text{d}]}} \; \bigg(\frac{D^{[\text{s}]} p^{[\text{t}]}}{D t}  - \alpha \; \frac{D^{[\text{f\,}]} p^{[\text{f\,}]}}{D t}   \bigg) - \frac{p^{[\text{t}]} - p^{[\text{f\,}]}}{\eta^{[\phi]}(1-\phi)} \; ,
\label{mass-sol} 
\end{align}
\end{fleqn}
\begin{fleqn}[\parindent]
\begin{align}
\nabla \cdot \boldsymbol{v}^{[\text{D}]} = \frac{\alpha}{K^{[\text{d}]}} \; \bigg(\frac{D^{[\text{s}]} p^{[\text{t}]}}{D t}  - \frac{1}{B} \; \frac{D^{[\text{f\,}]} p^{[\text{f\,}]}}{D t}   \bigg) + \frac{p^{[\text{t}]} - p^{[\text{f\,}]}}{\eta^{[\phi]}(1-\phi)}\; .
\label{mass-flu}   
\end{align}
\end{fleqn}

Superscripts [s], [f\,], [t], and [D] stand for \textit{solid}, \textit{fluid}, \textit{total} (solid and fluid) matrix, and \textit{Darcy}, respectively. $\boldsymbol{\underline{\underline{\sigma}}}$ is the Cauchy stress tensor, $\phi$ is the porosity, $\rho^{[\text{s}]}$ is the \textit{solid} density, $\boldsymbol{v}^{[\text{s}]}$ is the \textit{solid} velocity vector, $t$ is the time, $\boldsymbol{v}^{[\text{D}]}$ represents the \textit{Darcy flux} vector ($\boldsymbol{v}^{[\text{D}]}=\phi(\boldsymbol{v}^{[\text{f\,}]}-\boldsymbol{v}^{[\text{s}]})$), i.e., the volume flux relative to the moving solid matrix, $k^{[\phi]}$ is the porosity-dependent permeability, $\eta^{[\text{f} \,]}$ is the \textit{fluid} viscosity, $p^{[\text{t}]}$ and $p^{[\text{f\,}]}$ are the \textit{solid} and  \textit{fluid} pressure, respectively, $K^{[\text{d}]}$ is the \textit{drained} bulk modulus, $\eta^{[\phi]}$ is the effective compaction viscosity, $\boldsymbol{v}^{[\text{f\,}]}$ is the \textit{fluid} velocity vector, $\alpha$ is the Biot-Willis coefficient \citep{biot1957elastic}, $B$ is the Skempton coefficient \citep{skempton1960,bishop1973influence}, and $\boldsymbol{g}$ is the gravity vector. The total momentum (Eq. \ref{mom-sol+flu}) neglects differences in the solid and fluid accelerations \citep{gerya2007robust,petrini2020seismo}. \\
 
The porosity-dependent permeability is computed as 
 
\begin{fleqn}[\parindent]
\begin{align} 
k^{[\phi]}= k^* \,\bigg(\frac{\phi}{\phi^*}\bigg)^n \; ,
\label{perm-law}
\end{align} 
\end{fleqn}

where $k^*$ and $\phi^*$ are the reference permeability and reference porosity (Table \ref{table1}), respectively, and $n$ is an porosity-dependent exponent, which for natural pores is assumed to be 3 \citep{rice1992fault,connolly2000temperature}, and implies that permeability changes as a cube of increasing porosity. While different $k^*$ are tested in this study, $\phi^*$ is set to be 0.01 \citep[i.e. 1\%,][]{peacock2011high} and does not evolve with time. \\

The \textit{drained} bulk modulus ($K^{[\text{d}]}$), the Biot-Willis coefficient ($\alpha$), and the Skempton coefficient ($B$) are defined as

\begin{fleqn}[\parindent]
\begin{subequations}
\begin{align}
& K^{[\text{d}]} = (1-\phi) \bigg( \frac{1}{K^{[\phi]}} + \frac{1}{K^{[\text{s}]}} \bigg)^{-1} \; , \\[3pt]
&  \alpha = 1 - \frac{\beta^{[\text{s}]}}{\beta^{[\text{d}]}} \label{alpha}\; , \\[3pt]
&  B = \frac{\beta^{[\text{d}]}-\beta^{[\text{s}]}}{\beta^{[\text{d}]}-\beta^{[\text{s}]}+\phi(\beta^{[\text{f\,}]}-\beta^{[\text{s}]})} \label{Skempton} \; , \\[3pt]
& \beta^{[\text{d}]} = \frac{1}{\mu} + \frac{\beta^{[\text{s}]}}{1-\phi} \; , \label{dr-comp}
\end{align}
\end{subequations}
\end{fleqn}

where $K^{[\phi]}$ the \textit{effective} bulk modulus of pores, $K^{[\text{s}]}$ is the \textit{solid} bulk modulus, $\beta^{[\text{d}]}$ is the compressibility of the solid skeleton (i.e., \textit{drained} compressibility of the porous medium), $\beta^{[\text{s}]}$ is the compressibility of the \textit{solid} phase, and $\beta^{[\text{f\,}]}$ is the compressibility of the \textit{fluid} phase. \\

The stress is decomposed into a deviatoric and volumetric component:

\begin{fleqn}[\parindent]
\begin{align} 
\boldsymbol{\underline{\underline{\sigma}}} = \boldsymbol{\underline{\underline{\tau}}} - p^{[\text{t}]} \; \;
\boldsymbol{\underline{\underline{I}}}
\label{}
\end{align} 
\end{fleqn}

 where $\boldsymbol{\underline{\underline{\tau}}}$ is the deviatoric stress tensor, and $\boldsymbol{\underline{\underline{I}}}$ is the identity tensor. The total pressure ($p^{[\text{t}]}$) and density ($\rho^{[\text{t}]}$) are coupled with the porosity and computed from their respective fluid and solid matrix quantities

\begin{fleqn}[\parindent]
\begin{subequations}
\begin{align}
&  p^{[\text{t}]} = p^{[\text{f\,}]} \, \phi + p^{[\text{s}]}(1-\phi) \\[2pt]
&  \rho^{[\text{t}]} = \rho^{[\text{f\,}]} \, \phi + \rho^{[\text{s}]}(1-\phi)
\end{align}
\end{subequations}
\end{fleqn}

The effective bulk modulus of pores ($K^{[\phi]}$), as well as the effective compaction viscosity ($\eta^{[\phi]}$) are computed following \citet{yarushina2015compaction} and \citet{petrini2020seismo} as

\begin{fleqn}[\parindent]
\begin{subequations}
\begin{align}
&  K^{[\phi]}=\frac{2 \; m}{1+m} \; \frac{\mu}{\phi} \; , \label{effbulk} \\
&  \eta^{[\phi]}=\frac{2 \; m}{1+m} \; \frac{\eta_{\text{\,s}}}{\phi} \; , \label{effeta}
\end{align}
\end{subequations}
\end{fleqn}

where $m$ is a geometrical factor, $\mu$ is the shear modulus, and $\eta_{\text{\,s}}$ is the shear viscosity of the solid matrix. In this study $m$ takes a value of 1 (i.e., cylindrical pores), which reduces Eqs. (\ref{effbulk}) and (\ref{effeta}) to $\mu^* / \phi$ and $\eta_{\text{\,s}}/\phi$. For in-plane (mode-II) slip, the effective shear modulus, which is simply the effective elastic modulus for quasi-static plane-strain deformation \citep{rubin2007aftershock}, is equal to

\begin{fleqn}[\parindent]
\begin{align} 
\mu^*=\frac{\mu}{1-\nu} \; ,
\label{}
\end{align} 
\end{fleqn}

Where $\mu$ is the shear modulus, $\nu$ is Poisson's ratio. \\

The continuity equations of both solid (Eq. \ref{mass-sol}) and fluid (Eq. \ref{mass-flu}) contain poroelasticity terms consistent with Biot's theory \citep{biot1941general,gassmann1951elastic,yarushina2015compaction}, which allow for compressibility of the solid matrix and fluid as well as viscous and elastic (de)compaction of the interconnected porous space. Being Eqs. (\ref{mass-sol}) and (\ref{mass-flu}) fully coupled, convergence/divergence of the solid matrix occurs in response to local compaction/decompaction processes, forming a fully coupled hydro-mechanical system.

\begin{table}[t!]
\footnotesize
\centering
\begin{tabular}{ m{5cm}  m{3cm} m{3cm}}
\toprule
\textbf{Parameter} & Symbol & Value  \\
\midrule
$x$--domain                         &    $L_x$             &     100 km \\
$y$--domain                         &    $L_y$             &     20 km \\
Grid resolution      	           &    $\Delta_x$       &     250 m \\          
Shear modulus                    &    $\mu$             &      25 GPa \\
Bulk modulus 	                   &    $K$                &      50 GPa \\
Poisson ratio	                   &    $\nu$              &     0.25           \\
Total pressure                     &    $p^{[\text{t}]}$           &     40 MPa \\
Fluid pressure                     &    $p^{[\text{f\,}]}$         &    10 MPa \\
Solid density     	          &    $\rho^{[\text{s}]}$      &     2700 kg m$^{\text{-3}}$ \\
Fluid density     	                   &    $\rho^{[\text{f\,}]}$      &   1000 kg m$^{\text{-3}}$ \\
Solid compressibility     	  &    $\beta^{[\text{s}]}$      &     2.5 10$^{-11}$ 1/Pa \\
Fluid compressibility     	  &    $\beta^{[\text{f\,}]}$     &    4.0 10$^{-10}$ 1/Pa \\
Solid viscosity                     &    $\eta_{0}$                   &   10$^{23}$ Pa s \\
Fluid viscosity     	          &    $\eta^{[\text{f\,}]}$     &     10$^{-3}$ Pa s \\
Shear wave speed	           &    $c_s$       &     $\sqrt{\mu/\rho^{[\text{t}]}}$ \\
Gravity                                 &     $g$          &     9.81998 m s$^{\text{2}}$ \\
Reference friction                &     $f$        &     0.6 \\
Cohesion                             &     $c$        &     3.0 MPa \\
Fault width        	           &    $h$                &     100 m \\   
Critical nucleation size         &     $L_c$          &     7.3 km \\
Cohesive zone size             &     $\Lambda_0$    &2.5 m \\
Reference velocity               &     $V_0$            & 10$^{-9}$ m s$^{\text{-1}}$ \\
Reference strain rate           &     $\varepsilon_0$    &           $V_0/(2\,\Delta_x)$ \\
Reference porosity               &    $\phi^*$                 &          1$\%$ \\
Reference permeability        &    $k^*$                 &          10$^{-16}$ m$^2$ \\
Loading velocity (plate rate) &     $V_p$            & 2 $\cdot$ 10$^{-9}$ m s$^{\text{-1}}$ \\ [3pt]
\hline
\end{tabular}
\caption[]{\label{table1} Model parameters.}
\end{table}

\subsection{Visco-elasto-plastic rheology}
 The deviatoric deformation occurs as a combination of elastic, viscous, and plastic deformation. Components of the deviatoric strain-rate tensor are defined via spatial grandients of the velocity under the assumption of \textit{infinitesimal} strain as follows:

\begin{fleqn}[\parindent]
\begin{align} 
\dot{\varepsilon}_{ij}' = \frac{1}{2} \bigg(\frac{\partial v_i}{\partial x_j} + \frac{\partial v_j}{\partial x_i} \bigg) - \frac{1}{3} \mathrm{Tr} \; \dot{\varepsilon}_{ij} \: ,
\label{inf-strain}
\end{align} 
\end{fleqn}

 where $i$ and $j$ are coordinate indexes and $x_i$ and $x_j$ are spatial coordinates such that in 2-D we can define four tensor components. We assume that strain rate can be additively decomposed into its elastic, viscous, and plastic components as

\begin{fleqn}[\parindent]
\begin{align} 
\dot{\varepsilon}_{ij}' = [\, \dot{\varepsilon}_{ij}' \, ]_{\text{viscous}} + [\, \dot{\varepsilon}_{ij}' \, ]_{\text{elastic}} + [\, \dot{\varepsilon}_{ij}'  \, ]_{\text{plastic}} \: .
\label{strain_rate_total}
\end{align} 
\end{fleqn}

 The elastic part is defined as 

\begin{fleqn}[\parindent]
\begin{align} 
[\, \dot{\varepsilon}_{ij}' \, ]_{\text{elastic}}=\frac{1}{2 \mu} \; \frac{\tilde{D}}{\tilde{D} t} \: (\tau_{ij}) \: ,
\label{elastic}
\end{align} 
\end{fleqn}

 where $\mu$ is the shear modulus, while the differential operator $\tilde{D}/\tilde{D} t$ denotes the co-rotational time derivative --- that is, the local rates of change in the deviatoric stress with respect to translations, rotations, and/or deformation. \\

Viscous creep obeys Newton's law of viscous friction and is defined here via a single effective \textit{shear} viscosity according to

\begin{fleqn}[\parindent]
\begin{align} 
\eta_{\text{\,s}} = \frac{\tau_{II}^{\prime}}{2 \, \dot{\varepsilon}_{\text{II\,[v]}}^{\prime}} \:,
\label{eta_eff}
\end{align} 
\end{fleqn}

 where the $\tau_{II}^{\prime}$ and $\dot{\varepsilon}_{\text{II(v)}}^{\prime}$ denote the square root of the second invariant of the deviatoric stress and the viscous strain-rate tensors, respectively. The effective shear viscosity $\eta_{\text{\,s}}$ is then used to define the viscous contribution to the strain-rate tensor:

\begin{fleqn}[\parindent]
\begin{align} 
[\, \dot{\varepsilon}_{ij}' \, ]_{\text{viscous}} =  \frac{\tau_{ij}}{2 \, \eta_{\text{\,s}}} \: . 
\label{eta_eff}
\end{align} 
\end{fleqn}

Plastic (irreversible) deformation and frictional sliding is governed by a rate-dependent strength of a non-associated plasticity model \citep{yi2018simple}

\begin{fleqn}[\parindent]
\begin{align}
\tau_{II} = \tau_0 \, \bigg(\frac{\dot{\varepsilon}_{II \, [\text{p}]}'}{\varepsilon_{0}}\bigg)^{\gamma} \; , 
\label{syield}   
\end{align}
\end{fleqn}

where $\dot{\varepsilon}_{II \, [\text{p}]}'$ is the square root of the second invariant of the deviatoric plastic strain rate

\begin{fleqn}[\parindent]
\begin{align}
\dot{\varepsilon}_{II \, [\text{p}]}' = \varepsilon_{0} \, \bigg(\frac{\tau_{II}}{\tau_0}\bigg)^{1/\gamma}\; ,
\label{}   
\end{align}
\end{fleqn}

$\varepsilon_{0}$ is the reference strain rate, $\gamma$ is the rate-strengthening exponent, and $\tau_0$ represents the inclusion of the Drucker--Prager yield function defining the maximum allowable shear stress a rock can withstand \citep{prager1952soil}

\begin{fleqn}[\parindent]
\begin{align}
\tau_0 = c + f \, (p^{[\text{t}]} - p^{[\text{f\,}]}) 
\label{DruckerPrager}   
\end{align}
\end{fleqn}

where $c$ is the cohesional strength and $f$ is the reference, static friction coefficient. The slip rate $V(t)$ is then computed as the magnitude of the second invariant of deviatoric plastic strain rate $\dot{\varepsilon}_{\text{II(p)}}$ over the thickness of the fault zone $h$

\begin{fleqn}[\parindent]
\begin{align} 
V(t,x)=2 \; \dot{\varepsilon}_{II \, [\text{p}]}' \; h \; .
\label{slip_rate}
\end{align} 
\end{fleqn}

In our continuum model, plastic deformation is computed as volumetric strain over a finite thickness, and it is represented by a tensor. This means that plastic deformation can spontaneously localize anywhere \citep{velo2cycles}. Note that the rate-dependent strength (Eq. \ref{syield}) relies on the presence of $\gamma$, which mimic the positive direct effect in the rate- and state-dependent friction formulation \citep[e.g.,][]{scholz1998earthquakes}, a feature that has ample laboratory confirmation \citep{dieterich1979modeling,dieterich1981constitutive,ruina1983slip,marone1998laboratory}. On the other hand, our rate-dependent strength formulation does not require the presence of an evolutionary effect involving a decrease in friction. This is because the weakening mechanism in our fully coupled system is controlled by the evolution of pore-fluid pressure ($p^{[\text{f\,}]}$), which is encapsulated in $\tau_0$ (Eq. \ref{DruckerPrager}). \\

The non-associated plastic flow law is defined through the \textit{plastic flow potential} ($Q$), which reflects the amount of mechanical energy per unit volume that supports plastic deformation \citep{vermeer1998non}

\begin{fleqn}[\parindent]
\begin{align} 
Q = \tau_{II} - \text{sin}(\psi) \, (p^{[\text{t}]} - p^{[\text{f\,}]}) \, - \text{cos}(\psi) \: c \: .
\label{}
\end{align} 
\end{fleqn}

In this study, we assume the dilation angle $\psi$ to be zero, which implies that the volume does not change during plastic yielding ($Q = \tau_{II} - c $). Accordingly, the deviatoric plastic strain is defined as

\begin{fleqn}[\parindent]
\begin{align} 
[\, \dot{\varepsilon}_{ij}' \, ]_{\text{plastic}}=
\begin{cases}
    0      &   \rightarrow \tau_{II} < \tau_{y}\\
    \chi \frac{\partial Q}{\partial \tau_{ij}} = \chi \frac{\partial \tau_{ij}}{\partial \tau_{II}}  &  \rightarrow \tau_{II} = \tau_{y}\\  \end{cases}\: ,
\label{eps_pl}
\end{align} 
\end{fleqn}

 with $\chi$ indicating the \textit{plastic multiplier}, which serves to ensure that, if yielding occurs, the deviatoric stress will always satisfy the yield criteria ($\tau_{II} =\tau_{y}$). \\

The deviatoric components ($\tau_{ij}$) of the stress tensor $\boldsymbol{\underline{\underline{\sigma}}}$ in Eq. (\ref{mom-sol+flu}) are formulated from the visco-elasto-plastic constitutive relationships (Eq. \ref{strain_rate_total}) by using an implicit first-order finite-difference scheme in time in order to represent objective time derivatives of visco-elastic stresses \citep[e.g.,][]{Moresi2003476}

\begin{fleqn}[\parindent]
\begin{align} 
\tau_{ij} = 2 \, \eta_{\text{vp}}  \, Z \, \dot{\varepsilon}_{ij}' + \tau^{0}_{ij} \cdot (1-Z) \; ,
\label{full-stress}
\end{align} 
\end{fleqn}

where $Z$ is the visco-elasticity factor \citep{schmalholz2001spectral}

\begin{fleqn}[\parindent]
\begin{align} 
Z = \frac{\mu \, \Delta t}{\mu \, \Delta t + \eta_{\text{vp}}} \; ,
\label{visco-elasticity-factor}
\end{align} 
\end{fleqn}

  and $\eta_{\text{vp}}$ is the effective visco-plastic viscosity that characterizes the intensity of the plastic deformation

\begin{fleqn}[\parindent]
\begin{align} 
\eta_{\text{vp}} =
\begin{cases}
    \eta_{\text{\,m}}   &   \rightarrow \tau_{II} < \tau_{y}\\
    \eta_{\text{\,m}} \; \dfrac{\tau_{II}}{2 \, \eta_{\text{\,m}} \, \dot{\varepsilon}_{II \, \text{[p]}} \, + \tau_{II}}   
    &  \rightarrow \tau_{II} = \tau_{y}\\  \end{cases}\: ,
\label{}
\end{align} 
\end{fleqn}

where $\dot{\varepsilon}_{II \, \text{[p]}}$ is the square root of the second invariant of the deviatoric plastic strain rate, and $\eta_{\text{\,m}}$ is the \textit{matrix} viscosity \citep{katz2006dynamics}

\begin{fleqn}[\parindent]
\begin{align} 
\eta_{\text{\,m}} = \eta_{\text{\,s}} \; \text{e}^{(\lambda \; \phi)} \; ,
\label{}
\end{align} 
\end{fleqn}

in which $\lambda = -29 $ defines an experimentally derived porosity-weakening coefficient \citep{katz2006dynamics}. 
 
\subsection{Global visco-elasto-plastic Picard iterations}\label{picard-iter}
We accurately satisfy the plastic yielding condition at the Eulerian nodal points of the staggered grid by using global Picard iteration \citep{gerya2019introduction}. As a result, we discretize the rate-strengthening yielding condition (Eq. \ref{syield}) to satisfy the condition $\tau_{yield} = \tau_{II}$:

\begin{fleqn}[\parindent]
\begin{align}
\begin{split}
\tau_{yield \,(i,\, j)} = & \; c + f \, \, \frac{1}{4} \bigg((p_{(i,\, j)}^{[\text{t}]} - p_{(i,\, j)}^{[\text{f\,}]})+(p_{(i+1,\, j)}^{[\text{t}]} - p_{(i+1,\, j)}^{[\text{f\,}]}) \; .\,.\,.\\ 
& +(p_{(i,\, j+1)}^{[\text{t}]} - p_{(i,\, j+1)}^{[\text{f\,}]})+(p_{(i+1,\, j+1)}^{[\text{t}]} - p_{(i+1,\, j+1)}^{[\text{f\,}]})\bigg) \bigg(\frac{\dot{\varepsilon}_{[\text{p}]}'}{\varepsilon_{0}}\bigg)^{\gamma}  \; , 
\end{split}
\end{align}
\end{fleqn}

\begin{fleqn}[\parindent]
\begin{align}
\begin{split}
& \tau_{II \,(i,\, j)} = \; \Bigg({\tau_{xy \,(i,\, j)}}^2 + \frac{1}{2} \bigg(\frac{\tau_{xx \,(i,\, j)} + \tau_{xx \,(i+1,\, j)} + \tau_{xx \,(i,\, j+1)} + \tau_{xx \,(i+1,\, j+1)}}{4} \bigg)^2  \; .\,.\,. \\
& + \frac{1}{2} \bigg(\frac{\tau_{yy \,(i,\, j)} + \tau_{yy \,(i+1,\, j)} + \tau_{yy \,(i,\, j+1)} + \tau_{yy \,(i+1,\, j+1)}}{4} \bigg)^2  \; .\,.\,. \\
& + \frac{1}{2} \bigg(\frac{\tau_{xx \,(i,\, j)} - \tau_{yy \,(i,\, j)} + \tau_{xx \,(i+1,\, j)} - \tau_{yy \,(i+1,\, j)} + \tau_{xx \,(i,\, j+1)} - \tau_{yy \,(i,\, j+1)} + \tau_{xx \,(i+1,\, j+1)} - \tau_{yy \,(i+1,\, j+1)}}{4} \bigg)^2 \Bigg)^{1/2} 
\end{split}
\end{align}
\end{fleqn}

The Picard iteration is performed by repeating solution of Eqs (\ref{mom-sol+flu}$-$\ref{mass-flu}) and evaluating $[\, \tau_{II} \, ]_{\text{elastic}}$ at each basic node as

\begin{fleqn}[\parindent]
\begin{align} 
[\, \tau_{II} \, ]_{\text{elastic}} = \tau_{II} \bigg(\frac{\mu^* \, \Delta t + \eta_{\text{\,vp}}}{\eta_{\text{\,vp}}} \bigg) \; ,
\label{}
\end{align} 
\end{fleqn}

Where $\eta_{\text{\,vp}}$ is the current local value of the effective visco-plastic viscosity of the solid matrix, which defines the relationship between the deviatoric stress and the deviatoric strain rate for a viscous rheology. When $[\, \tau_{II} \, ]_{\text{elastic}} > \tau_{yield}$, we compute the new visco-plastic viscosity at the node as

\begin{fleqn}[\parindent]
\begin{align} 
\eta_{\text{vp}} = \mu^* \, \Delta t \, \frac{\tau_{yield}}{[\, \tau_{II} \, ]_{\text{elastic}} - \tau_{yield}}\, .
\label{visco-plastic-visc}
\end{align} 
\end{fleqn}

If $\eta_{\text{vp}} <  \eta_{\text{\,s}}$, the new visco-plastic viscosity from Eq. (\ref{visco-plastic-visc}) is used for the next iteration. The global plastic yielding condition error is then computed on the nodal points 

\begin{fleqn}[\parindent]
\begin{align} 
\Delta \tau_{yield} = \frac{1}{N} \sqrt{\sum_{i,\, j} \Big(\tau_{II \,(i,\, j)}-\tau_{yield \,(i,\, j)}\Big)^2}
\label{}
\end{align} 
\end{fleqn}

where $N$ is the cumulative number of nodal points at which either the previously computed or new value of $\eta_{\text{vp}}$ satisfies the condition $\eta_{\text{vp}} <  \eta_{\text{\,s}}$. The stopping criteria of the Picard iteration is reached when $\Delta \tau_{yield}$ has decreased below a desirable level of tolerance ($\delta_{err}$)

\begin{fleqn}[\parindent]
\begin{align} 
\Delta \tau_{yield} \le \delta_{err} \,,
\label{}
\end{align} 
\end{fleqn}

where $\delta_{err}=10 \; \text{Pa}$. \\

An important mechanism for fluid pressurization is the feedback between visco-plastic viscosity and \textit{compaction viscosity} (often called bulk viscosity), which characterizes solid matrix resistance to reversible (i.e., elastic, $\beta^{[\text{d}]}$) and irreversible (i.e., visco-plastic, $\eta^{[\phi]}$) pore compaction/decompaction. When plastic yielding occurs, we first compute the plastic viscosity ($\eta_{\text{\,p}} $) at the basic nodes (b) as

\begin{fleqn}[\parindent]
\begin{align} 
\eta_{\text{\,p\,(b)}} = \frac{1}{1/\eta_{\text{\,vp\,(b)}} - 1/\eta_{\text{\,m\,(b)}}}\,.
\label{}
\end{align} 
\end{fleqn}

We then compute the new plastic viscosity at the center of cells (i.e., pressure nodes, p) by using the harmonic average of the plastic viscosity values from the four surrounding basic nodes (b) \citep{gerya2019introduction}

\begin{fleqn}[\parindent]
\begin{align} 
\eta_{\text{\,p\,(p)\,}(i,j)} = \frac{4}{1/\eta_{\text{\,p(b)}(i,j)} + 1/\eta_{\text{\,p(b)}(i-1,j)} + 1/\eta_{\text{\,p(b)}(i,j-1)} + 1/\eta_{\text{\,p(b)}(i-1,j-1)}} \,.
\label{}
\end{align} 
\end{fleqn}

Finally we compute the new compaction viscosity at the pressure nodes as 

\begin{fleqn}[\parindent]
\begin{align} 
\eta_{\text{(p)} (i,j)}^{[\phi]} = \frac{1}{\phi \; (1/\eta_{0 \text{(p)} (i,j)} + 1/\eta_{\text{\,p\,(p)}(i,j)})}\,,
\label{}
\end{align} 
\end{fleqn}

and the new visco-plastic viscosity at the pressure nodes as

\begin{fleqn}[\parindent]
\begin{align} 
\eta_{\text{\,vp\,(p)}(i,j)} = \frac{1}{1/\eta_{0 \text{(p)} (i,j)} + 1/\eta_{\text{\,m\,(p)}(i,j)}}\,,
\label{}
\end{align} 
\end{fleqn}

where $\eta_{0}$ is the reference solid viscosity on the pressure nodes (Table \ref{table1}).

\subsection{Adaptive time-stepping}
To resolve different time-scales --- ranging from long-term (slow) deformation to periodic (fast) slip events --- a variable time stepping is required during computation. To accomplish this, we require that the time step is the minimum of the time steps needed to resolve (1) the slip acceleration on the fault ($\Delta t_{\text{s}}$), (2) the displacement per grid cell ($\Delta t_{\text{d}}$), and (3) the visco-elasto-plastic time step ($\Delta t_{\text{vep}}$) as

\begin{fleqn}[\parindent]
\begin{align} 
\Delta t= \text{min:} \, \Big\{ \Delta t_{\text{s}}  \, , \, \Delta t_{\text{d}}  \, , \, \Delta t_{\text{vep}}  \Big\}\: .
\label{2-DT}
\end{align} 
\end{fleqn}

The first time step requires that the slip acceleration per time step is limited to a fraction of the grid size as

\begin{fleqn}[\parindent]
\begin{align} 
\Delta t_{\text{s}} = \frac{\Delta x \; \, \delta_d}{V_{\text{max}}}   \: ,
\label{}
\end{align} 
\end{fleqn}

where $\delta_d= 10^{-5}$ defines the maximum grid fraction. Similarly, we require that the displacement per time step is limited by the same fraction of the grid size

\begin{fleqn}[\parindent]
\begin{align} 
\Delta t_{\text{d}} = \delta_d \; \; \text{min:} \Bigg\{ \bigg| \frac{\Delta x}{V_x} \bigg| \, , \, \bigg| \frac{\Delta y}{V_y} \bigg| \Bigg\}  \: .
\label{}
\end{align} 
\end{fleqn}

Based on a previous study \citep{herrendorfer2018modeling}, we apply a visco-elasto-plastic time step to capture the relaxation time scale $\eta_{\text{vp}}/ \mu^*$ by a fraction $\xi$

\begin{fleqn}[\parindent]
\begin{align} 
\Delta t_{\text{vep}} = \xi \; \frac{\eta_{\text{vep}}}{\mu^*}  \: ,
\label{}
\end{align} 
\end{fleqn}

where $\xi = 0.2 $. Notably, the visco-elasto-plastic relaxation time step combines both the visco-elastic relaxation $\eta_{\text{vep}}/ \mu^*$ and the elasto-plastic relaxation time scale 

\begin{fleqn}[\parindent]
\begin{align} 
\frac{\tau_{II}}{\mu^* \chi \;} = \frac{\tau_{II}}{2 \; \mu^* \; [\, \dot{\varepsilon}_{ij}' \, ]_{\text{plastic}}} \: .
\label{}
\end{align} 
\end{fleqn}

As a result, $\Delta t_{\text{vep}}$ provides a constraint on the final time step when viscous deformation is dominant, thus ensuring the resolution of the relaxation time. Importantly, we do not apply any minimum time step cutoff. This is in contrast to previous studies \citep[e.g.,][]{lapusta2009three}, in which the time step is only adapted until it reaches $\Delta t_{\text{min}} = (1/3) \, \Delta_x / \, c_s \: ,$ where $c_s$ is the shear wave speed (see Table \ref{table1}). We do not adopt a minimum time step cutoff because during the propagation of fully dynamic ruptures the actual time step can decreases below the $\Delta t_{\text{min}}$ threshold. As a result, applying a cutoff would reduces the convergence rate during Picard iterations (Section \ref{picard-iter}).

\subsection{Spatial discretization}
We adopt a two-dimensional (2-D), staggered-grid finite difference--marker-in-cell (SFD-MIC) method, in which the grid is comprised of quadrilaterals cells parallel to the underlying coordinate basis vectors \citep{gerya2007robust}. Discretization of governing equations is given in \ref{discretization}. The numerical method is developed for an in-plane problem using a fully-staggered grid, which means that physical variables are defined at different geometric points (Fig. \ref{fig-grid}). In order to model multiple rock types and permit these rocks to undergo large displacement, as it often occurs on geologic faults, the Eulerian spatial discretization is combined with a Lagrangian marker-in-cell advection scheme. A set of Lagrangian particles (or material points) advects through the Eulerian domain used to discretize the velocity, pressure, and stress field. The Lagrangian particles are used to assign rock properties (e.g., density, viscosity, shear modulus, and hydro-mechanical properties), and, during the simulation, they store deviatoric stresses, pressure, and velocities for evaluating the respective time derivatives. A 4-th Runge-Kutta scheme is then used to advect markers to their new position according to the velocity field, whereas stresses are rotated according to the vorticity field. Lastly, a standard distance-dependent bi-linear interpolation scheme is used to interpolate physical parameters from the Lagrangian markers to the staggered Eulerian nodes at the beginning of each time step and update marker values at the end of each time step. \\

\subsection{Model setup and boundary conditions}\label{modelsetup}
\begin{figure}[h!]
\makebox[\textwidth][c]{\includegraphics[width=0.75\textwidth]{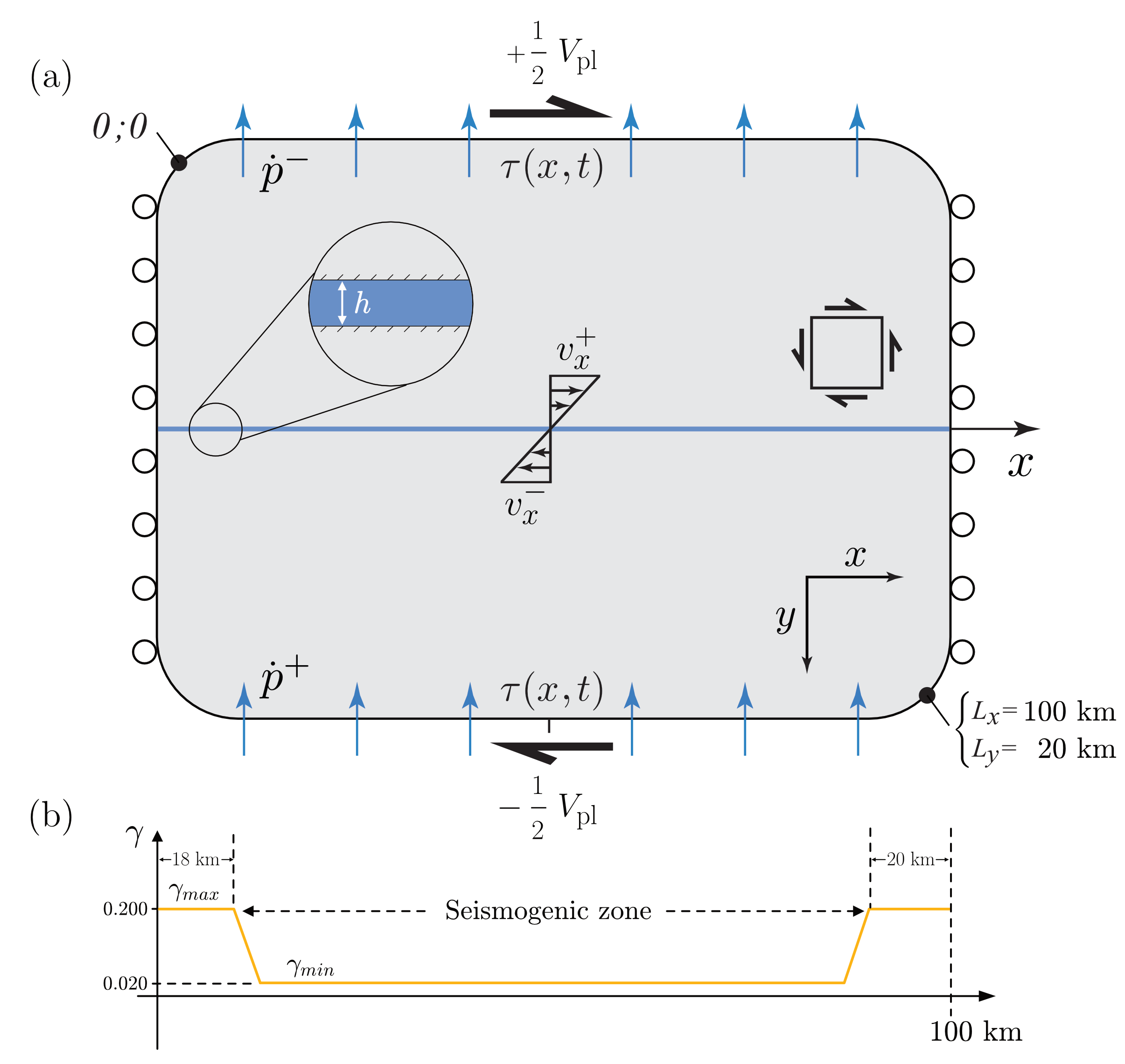}}
\caption{Illustration of the model setup. In-plane shear zone embedded in a homogeneous poro-visco-elasto-plastic media, with the shear layer undergoing uniform strain and a relative displacement. Boundary conditions include: free slip (small rollers) at the left and right boundaries, whereas tangential displacement (black arrows) is prescribed at the top and bottom boundaries. Blue arrows at the top and bottom boundaries indicate the inflow ($\dot{p}^+$) and outflow ($\dot{p}^-$) boundary conditions of fluid. The seismogenic region with velocity-weakening properties is surrounded by velocity-strengthening regions.}
\label{model_setup}
\end{figure}

 The 2-D model setup consists of a dextral strike-slip fault (Fig. \ref{model_setup}). We consider a mode-II plane-strain shear motion along a planar interface embedded in a poro-visco-elasto-plastic medium (Fig. \ref{model_setup}). We utilize an $x-y $ Cartesian coordinate system with $y=0$ being the sliding interface. This simple setup is designed to enable comparison to both analytical benchmarks for numerical solution and classical earthquake cycle simulations and theoretical estimates of the critical length scales during nucleation and propagation of dynamic ruptures \citep[e.g.,][]{lapusta2000elastodynamic}. Parameters for the reference model setup are given in Table \ref{table1}.\\
 
The system of equations over the domain $\Omega = [0 , L_x]\times[0 , L_y]$ has a unique solution given appropriate boundary conditions for the velocity and pressure (Fig. \ref{model_setup}). For the specifics of the model setup presented in this study, the conditions on the left ($x=0$) and right ($x=L_x$) boundaries are defined \textit{free slip} as follow:
 
\begin{fleqn}[\parindent]
\begin{subequations}
\begin{align}
& \frac{\partial v_y}{\partial x} = 0 \bigg|_{x=0\; \;} \; \; \; , \; v_x = 0 \bigg|_{x=0 \; \;}  \\
& \frac{\partial v_y}{\partial x} = 0   \bigg|_{x=L_x}  \; \; \, , \; v_x = 0 \bigg|_{x=L_x}  
\end{align}
\end{subequations}
\end{fleqn}

A \textit{free slip} condition requires that the normal velocity component on the boundary is zero and the other component do not change across the boundary. On the other hand, the fault is loaded from the upper ($y=0$) and lower boundaries ($y=L_y$) by applying Dirichlet boundary conditions for horizontal velocity $v_x$ and zero velocity on the vertical component $v_y$ (Fig. \ref{model_setup}):

\begin{fleqn}[\parindent]
\begin{subequations}
\begin{align}
&  v_x = + \frac{1}{2} \; V_{\text{pl}} \, \bigg|_{y=0} \; \; \;  \;  , \; \; v_y = 0 \, \bigg|_{y=0} \; \; \; ;\\[2pt]
&  v_x = - \frac{1}{2} \; V_{\text{pl}} \, \bigg|_{y=L_y} \; \; , \; \; v_y = 0 \, \bigg|_{y=L_y} \; ;
\end{align}
\end{subequations}
\end{fleqn}

 where $V_{\text{pl}}=6.3$ cm yr$^{-1}$ corresponds to the long-term loading rate. \\

 Pressure boundary conditions have to be defined for both total ($p^{[\text{t}]}$) and fluid ($p^{[\text{f\,}]}$) pressure. Since the solid mass (Eq. \ref{mass-sol}) and fluid mass (Eq. \ref{mass-flu}) are intrinsically coupled, we first initialize the fluid pressure with a constant confining pressure ($p_{\text{conf}}$) and then we apply a constant effective pressure ($p_{\text{eff}}$) on the top and lower boundary: 

\begin{fleqn}[\parindent]
\begin{align}
\boldsymbol{p}^{[\text{t}]} - \boldsymbol{p}^{[\text{f\,}]} = p_{\text{eff}} \; \bigg|_{y=0 \; \; , \; y=L_y} 
\end{align}
\end{fleqn}

Fluid momentum (Eq. \ref{mom-flu}) is solved assuming \textit{free slip} condition for the left and right boundaries

\begin{fleqn}[\parindent]
\begin{subequations}
\begin{align}
& \frac{\partial \boldsymbol{v_y}^{[\text{f\,}]}}{\partial x} = 0 \;  \bigg|_{x=0} \; \; \; , \; \boldsymbol{v_x}^{[\text{f\,}]}= 0  \;  \bigg|_{x=0} \\
& \frac{\partial \boldsymbol{v_y}^{[\text{f\,}]}}{\partial x} = 0 \; \bigg|_{x=L_x}  \; , \; \boldsymbol{v_x}^{[\text{f\,}]} = 0  \; \bigg|_{x=L_x}  
\end{align}
\end{subequations}
\end{fleqn}

whereas an inward ($\dot{p}^+$) and outward ($\dot{p}^-$) fluid flux is defined on the top and lower boundary in the form of Neumann boundary conditions

\begin{fleqn}[\parindent]
\begin{subequations}
\begin{align}
& \frac{\partial \boldsymbol{v_y}^{[\text{f\,}]}}{\partial y} = \dot{p}^- \; \bigg|_{y=0} \; \; \; , \; \boldsymbol{v_x}^{[\text{f\,}]}= 0  \; \bigg|_{y=0} \\
& \frac{\partial \boldsymbol{v_y}^{[\text{f\,}]}}{\partial y} = \dot{p}^+ \; \bigg|_{y=L_y}  \; , \; \boldsymbol{v_x}^{[\text{f\,}]} = 0 \; \bigg|_{y=L_y} . 
\end{align}
\end{subequations}
\end{fleqn}

\section{Results}\label{results}

\subsection{Fluid injection benchmarks}
\begin{figure}[h!]
\makebox[\textwidth][c]{\includegraphics[width=1.0\textwidth]{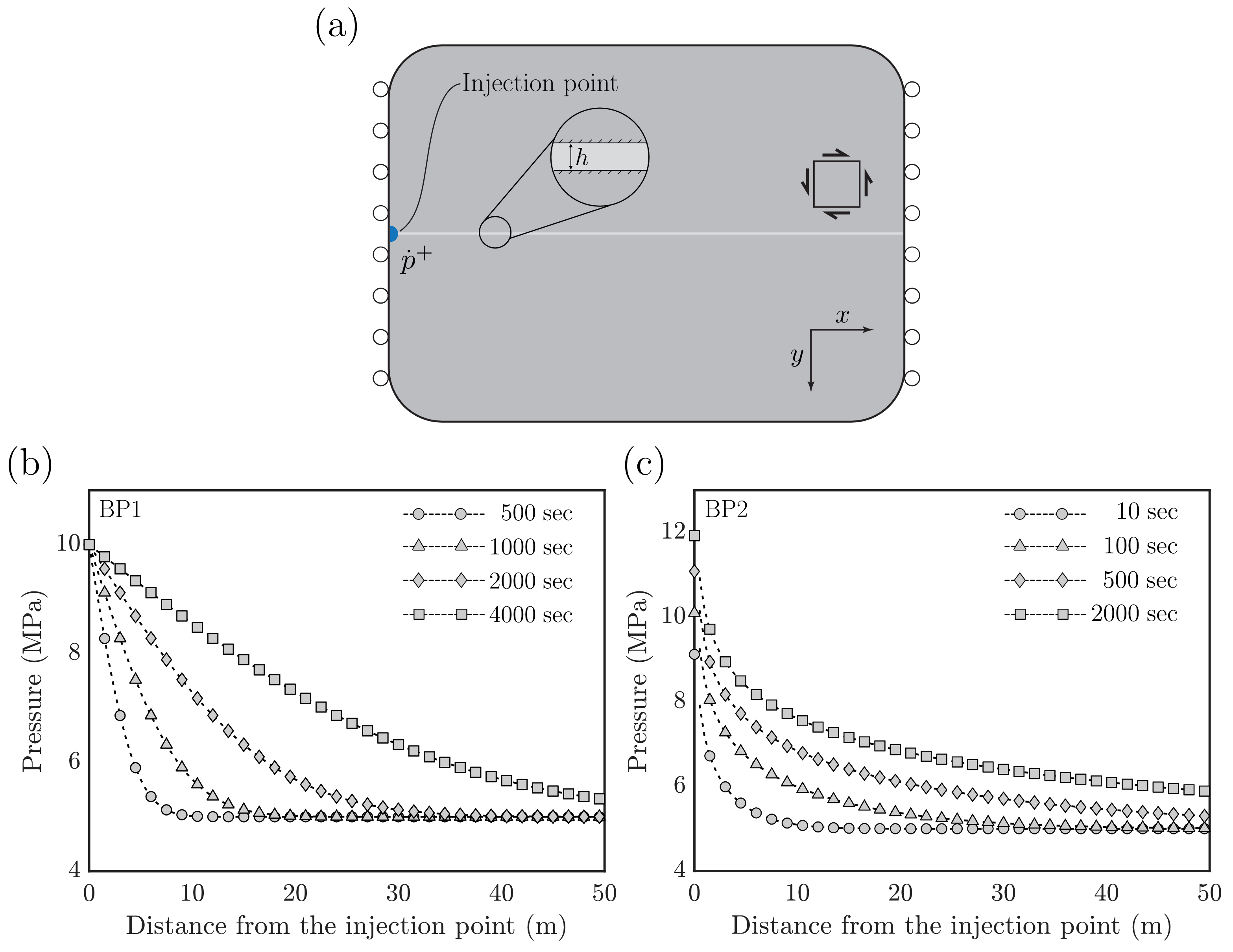}}
\caption{Modified model setup with a source point of fluid injection and results from analytical benchmarks of pore-fluid pressure diffusion. \textbf{(a)} The fault is embedded within a thin poro-visco-elasto-plastic layer of finite width, assumed to be much more permeable than surroundings. Fluid is injected at $x=0$ and diffuses along the fault with a diffusion timescale of $\sqrt{4 \, \alpha \, t}$. In both experiments, $p_0^{[\text{f\,}]}=5$ MPa, $\phi=1\%$, $\eta^{[\text{f\,}]}=10^{-3}$ Pa s, $k^{[\phi]}=10^{-17}$ m$^2$ and $h=1$ m. \textbf{(b)} First benchmark problem (BP1), which simulates fluid injection as a source of constant fluid pressure $\Delta p^{[\text{f\,}]}=5$ MPa. The black dashed lines indicate the analytical solution following Eq. (\ref{BP1}), whereas grey markers indicate the numerical results. \textbf{(c)} Second benchmark problem (BP2), which simulates fluid injection at a constant volume rate of $Q_w=2.8$ l/min. The black dashed lines indicate the analytical solution following Eq. (\ref{BP2}), whereas grey markers indicate the numerical results.}
\label{fig-benchmark}
\end{figure}

To test the robustness of the code in a broad range of applications relevant to solid-fluid interaction, in this section we compare our modeling results against analytical benchmarks of pore-fluid pressure diffusion. In particular, we consider two benchmark exercises (BP1 and BP2) from a set of quasi-static problems in a 2-D domain with a 1-D fault subjected to perturbations of fluid injection and along-fault pore-fluid diffusion. In order to verify these two benchmarks, we modify the model setup given in Section \ref{modelsetup} by including an injection point on the left boundary (Fig. \ref{fig-benchmark}a). \\

For the first benchmark problem (BP1), injection is modeled as a source of constant fluid pressure ($\Delta p^{[\text{f\,}]}$) at $x_{inj}=x_{(0)}$. The resulting pore-fluid pressure diffusion along the interface can be solved analytically as \citep[e.g.,][]{viesca2021self}

\begin{fleqn}[\parindent]
\begin{align} 
p^{[\text{f\,}]}(x,t) = p_0^{[\text{f\,}]} + \Delta p^{[\text{f\,}]} \; \cdot \text{erfc} \bigg( \frac{|x|}{l_d (t)} \bigg) \; ,
\label{BP1}
\end{align} 
\end{fleqn}

where erfc is the complementary error function, $l_d(t) = \sqrt{4 \, \alpha \, t}$ is the diffusion lengthscale, and $\alpha$ is the constant hydraulic diffusivity

\begin{fleqn}[\parindent]
\begin{align} 
\alpha = \frac{k^{[\phi]}}{\eta^{[\text{f\,}]} \; \beta^*}\ \; ,
\label{}
\end{align} 
\end{fleqn}

where $\eta^{[\text{f\,}]}$ is the viscosity of the permeating fluid and $\beta^*$  is a storage coefficient reflecting the compressibility of the fluid and porous matrix.
On the other hand, the second benchmark problem (BP2) assumes a constant volume rate injection from a point source ($x_{inj}=x_{(0)}$), and the flow solution is given by \citet{carslaw1959conduction}:

\begin{fleqn}[\parindent]
\begin{align} 
p(x,t)-p_0=\frac{Q_w \; \eta^{[\text{f\,}]}}{4 \; \pi \; k^{[\phi]} \; h} \; \; E_1 \Bigg( \frac{\sqrt{(x-x_{inj})^2}}{4\; \alpha \; t} \; \Bigg) \; ,
\label{BP2}
\end{align} 
\end{fleqn}

where $Q_w$ is the constant injection volume rate, $h$ is the fault thickness, and E1 is the expo- nential integral function. Note that the fault hydraulic transmissivity represents the product $k^{[\phi]} \, h$. \\

For both benchmarks BP1 and BP2, fluid is injected into the fault and it is constrained to flow only within its hydraulic aperture, a scenario that would occur when the permeability of the host rock is negligible relative to the fault itself (i.e. no fluid leak-off). Also, the fault is assumed \textit{marginally pressurized} --- that is, $\Delta p^{[\text{f\,}]}$ is always below the total pressure (in order to avoid hydraulic fracturing). Figure \ref{fig-benchmark} shows the results from the BP1 (Fig. \ref{fig-benchmark}b) and BP2 (Fig. \ref{fig-benchmark}c) and demonstrates the high accuracy of the numerical solution, which overlaps with the analytical solutions at different time stages. It is worth noting that the exponential integral function in Eq. (\ref{BP2}) is singular at the injection point $x = x_{inj}$ (Fig. \ref{fig-benchmark}c). This means that, for the case of injection experiments at constant volume rate, the fluid pressure at the origin is unbounded.

\subsection{Earthquake cycles}
\begin{figure}[h!]
\makebox[\textwidth][c]{\includegraphics[width=1.0\textwidth]{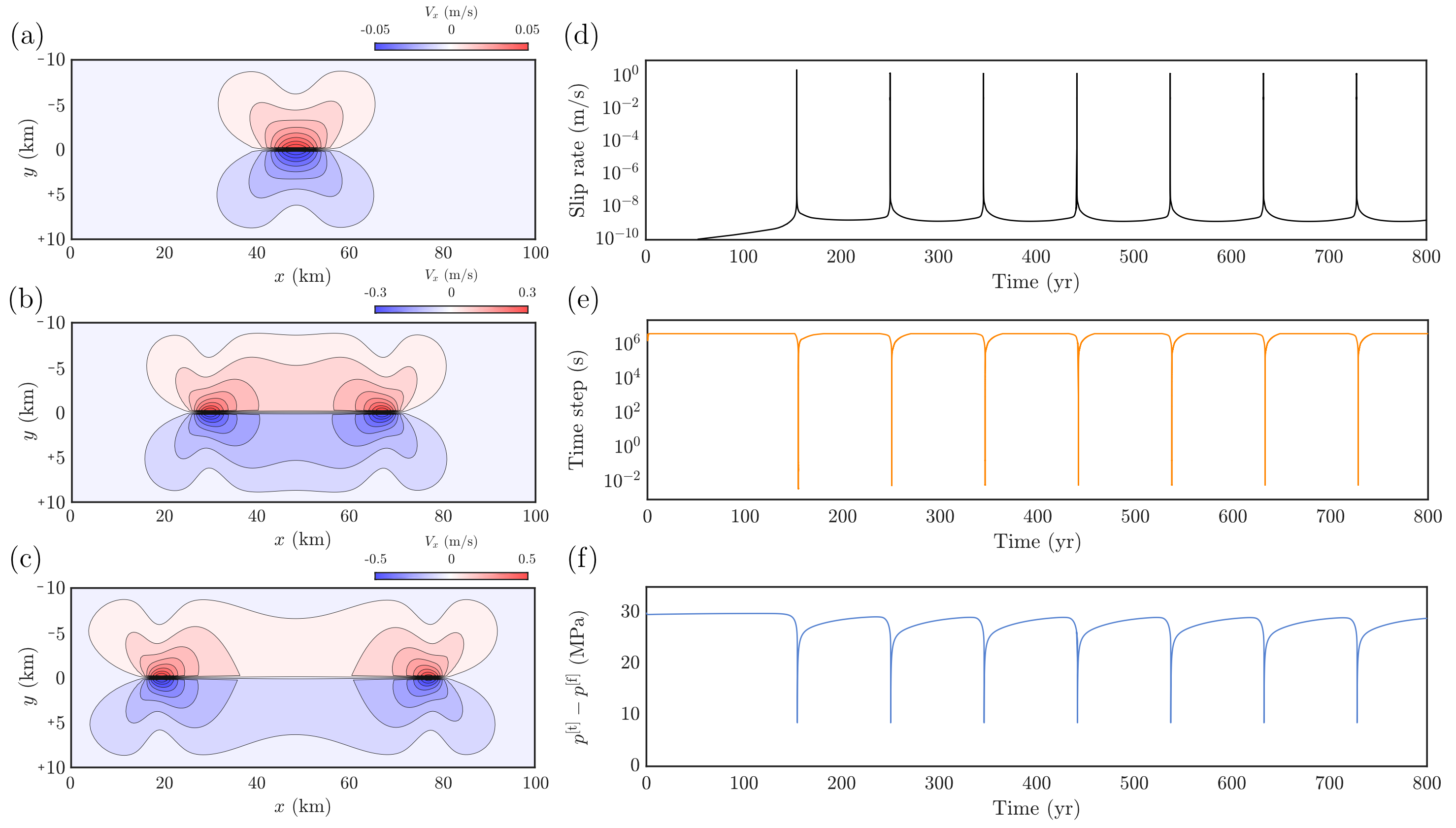}}
\caption{Spatial horizontal velocity and temporal evolution of earthquake cycles. Snapshots of the spatial horizontal velocity ($v_x$) distribution during the phases of (\textbf{a}) nucleation, and (\textbf{b--c}) dynamic propagation. The response of the fault to tectonic loading is characterized by long inter-seismic periods of quasi-static deformation, in which the seismogenic zone is practically locked, whereas lateral segments of the fault zone steadily creeps at rates comparable to the plate convergence rate. Note that the color scale of $v_x$ velocities in (\textbf{a--c}) are not the same. Temporal evolution of (\textbf{d}) maximum slip velocity (\textbf{e}) time-step, and (\textbf{f}) effective pressure on the fault. The time step is set to be inversely proportional to slip velocity, which results in large time steps in the inter-seismic periods, and small time steps (of the order of milliseconds) to simulate the evolution of each dynamic rupture.}
\label{eq-cycles}
\end{figure}

The response of the fault to slow tectonic loading is characterized by long periods of quasi-static deformation followed by short, and fast, slip events (Fig. \ref{eq-cycles}a--c), in which the slip rate on the fault accelerates from $\sim$cm yr$^{-1}$ to $\sim$m s$^{-1}$ (Fig. \ref{eq-cycles}d). Since the time step is set to be inversely proportional to slip velocity, relatively large time steps --- of the order of a fraction of a year --- are used in the inter-seismic period, while small time steps of the order of milliseconds are used to simulate the evolution of each dynamic rupture (Fig. \ref{eq-cycles}e). Similarly, the effective pressure  ($\boldsymbol{p}^{[\text{t}]} - \boldsymbol{p}^{[\text{f\,}]}$) drops by $\sim$20 MPa as a result of the poroelastic response of the fault zone during fast shearing (Fig. \ref{eq-cycles}f). \\

\begin{figure}[h!]
\makebox[\textwidth][c]{\includegraphics[width=1.0\textwidth]{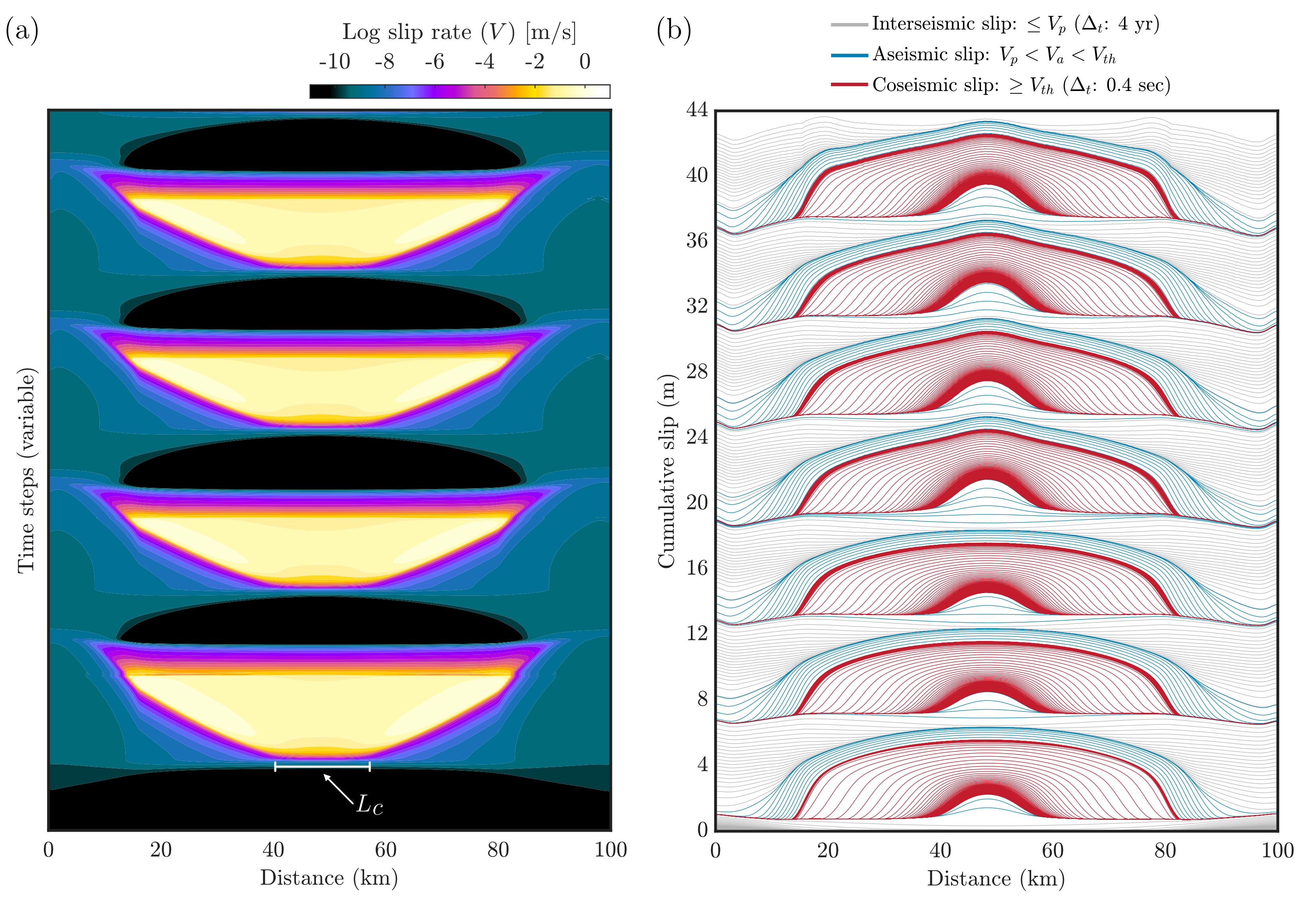}}
\caption{(\textbf{a}) Long-term histories of slip rate on the fault for the reference model. Note that, in order to visualize the evolution of dynamic ruptures, the $y$-axis displays the time step. The simulation period is 500 years, which start with a $\sim$150 years spin-up period. $L_c$ indicates the critical nucleation size. Slip velocity is plotted on the logarithmic scale. (\textbf{b}) Cumulative slip on the fault after multiple events over a period of 800 years. Red lines indicate the co-seismic slip every 0.4 sec when the maximum slip velocity exceeds the threshold of 8.1 cm s$^{-1}$ (see Eq. \ref{Vth}), blue lines illustrate post-seismic slip transients with slip rate in between the loading rate ($V_p$) and the co-seismic slip velocity threshold ($V_{th}$). Gray lines (every 4 years) illustrate the inter-seismic (aseismic) behavior of the fault with slip rates equal or lower than the loading rate.}
\label{fig-slip-history}
\end{figure}

Slip history on the fault indicate that, when assuming homogeneous properties, regular cycles of complete ruptures arises from the numerical experiments (Fig. \ref{fig-slip-history}). In the early stages of the simulations, the imposed loading rate increases stress linearly with time, whereas slip rate increases exponentially with time on the left and right side of the fault, i.e., where $\gamma$ is high (Fig. \ref{model_setup}b). As a result, while the two lateral segments of the fault steadily creep at rates comparable to the imposed loading rate (Fig. \ref{fig-slip-history}a), the seismogenic zone remains locked during inter-seismic periods and stress concentrates at the transition between high and low $\gamma$. In the later stages of the inter-seismic periods, the rate-strengthening behaviour allows creep to penetrate from both the left and the right segments towards the center of the fault, thus leading to a mechanical erosion of the locked patch. Due to a relatively large critical nucleation size ($L_c$), seismic events nucleate at the center of the seismogenic zone and the resulting dynamic ruptures typically propagate in a bilateral and symmetric fashion (Fig. \ref{fig-slip-history}a). For the reference model, the average slip rate is 0.82 m s$^{-1}$, whereas the average rupture speed is $\sim$2.35 km s$^{-1}$. \\

Evolution of the cumulative slip over multiple earthquake cycles is displayed in Figure \ref{fig-slip-history}b, with the accumulation of slip on the fault during inter-seismic periods represented by gray lines plotted every 4 years, whereas red lines display cumulative slip every 0.4 s when the maximum slip rate on the fault exceeds the following velocity threshold ($V_{th}$)

\begin{figure}[h!]
\makebox[\textwidth][c]{\includegraphics[width=1.0\textwidth]{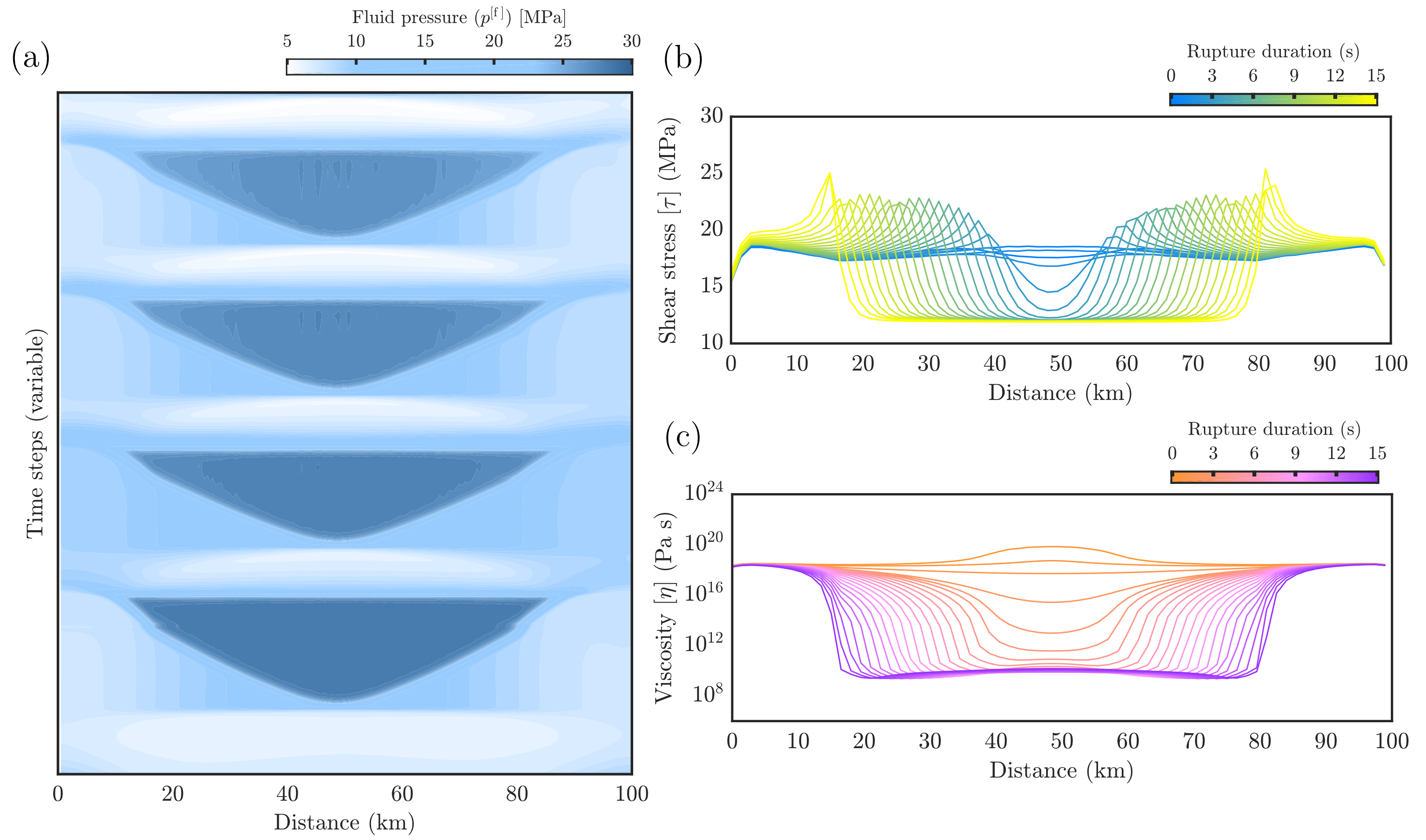}}
\caption{(\textbf{a}) Long-term histories of fluid pressure on the fault for the reference model. The $y$-axis displays the time step, which varies by several orders of magnitude. The simulation period is 500 years, which start with a $\sim$150 years spin-up period. Spatiotemporal evolution of (\textbf{b}) shear stress and (\textbf{c}) viscosity on the fault during a representative dynamic rupture. Colorbars indicate the temporal profile on the fault from the beginning of the dynamic rupture. Seismic events cause a sudden increase in pore-fluid pressure, a dynamic stress change, and a drop in viscosity which promotes co-seismic weakening.}
\label{pf-tau-eta}
\end{figure}

\begin{fleqn}[\parindent]
\begin{align} 
V_{th} = \frac{2 \; \gamma \; p_{\text{eff}} \; c_s}{\mu} \: ,
\label{Vth}
\end{align} 
\end{fleqn}

which, according to our parameters, yields to a threshold of 8.1 cm s$^{-1}$. When the fault is experiencing a seismic event, the cumulative slip on the fault indicates that a complete fault rupture produces $\sim$5 m of slip, which is followed by a transient post-seismic deformation of $\sim$0.8--1 m (blue lines; Fig. \ref{fig-slip-history}b). Our models show that dynamic ruptures occur due to an abrupt increase of pore-fluid pressure within the fault zone (Fig. \ref{pf-tau-eta}a). The increase in pore-fluid pressure is simultaneously coupled to a rapid decrease in shear strength (Fig. \ref{pf-tau-eta}b) and both shear and compaction viscosity (Fig. \ref{pf-tau-eta}c), which are self-sustained by a localized strain rate. When an event begins to propagate dynamically, it produces a shear stress breakdown from its maximum static value to a low, dynamic value (Fig. \ref{pf-tau-eta}b). Notably, at co-seismic slip rates, the effective shear viscosity drops by roughly 10 orders of magnitude (Fig. \ref{pf-tau-eta}c), whereas static stress change immediately after the event results in a stress concentration at the transition between the lateral creeping segments and the seismogenic zone. \\

\begin{figure}[h!]
\makebox[\textwidth][c]{\includegraphics[width=1\textwidth]{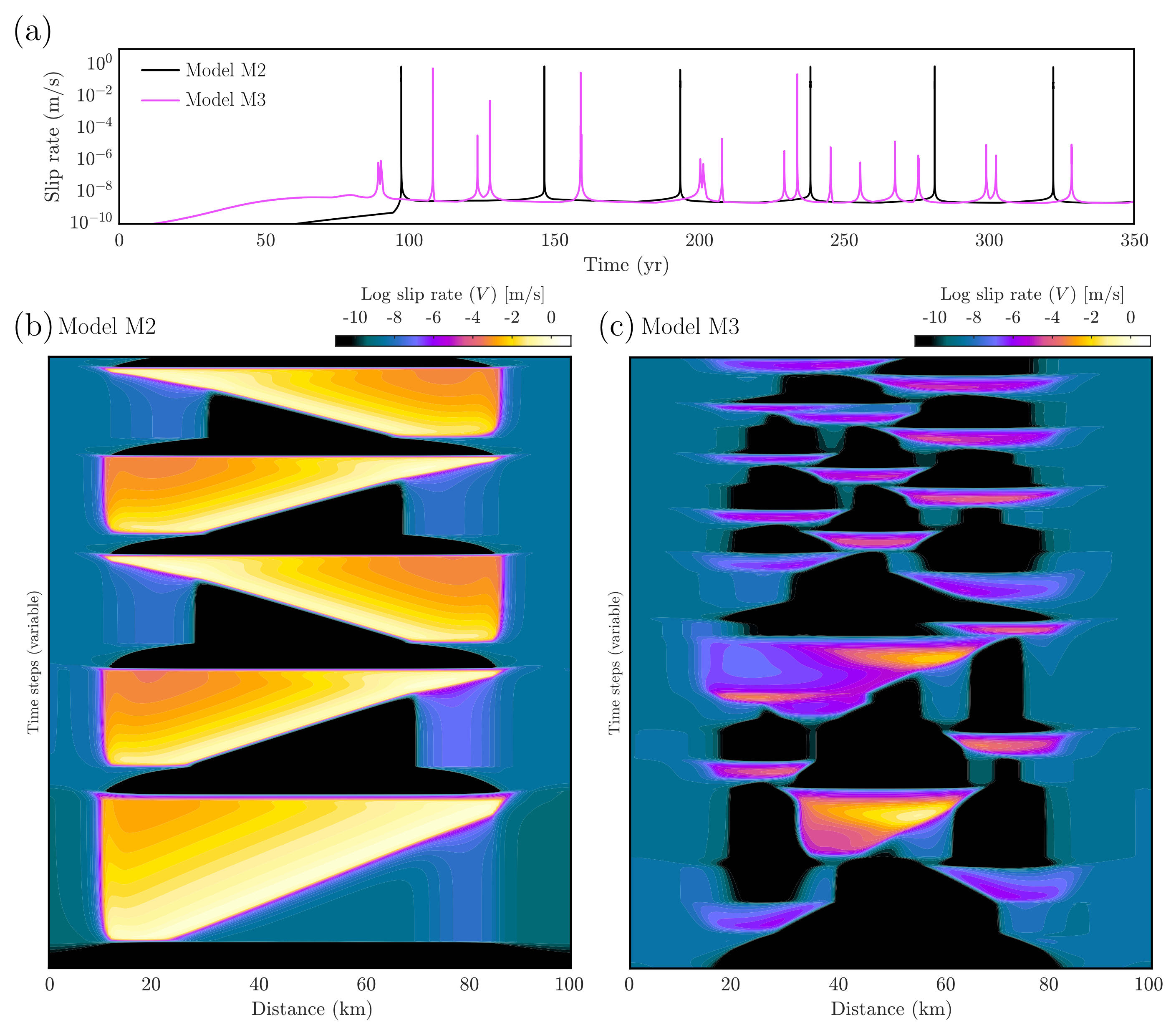}}
\caption{Overview of the results for the simulations with different effective pressure. (\textbf{a}) Temporal evolution of maximum slip velocity on the fault for models M2 and M3. (\textbf{b}) Model M2 ($p_{\text{eff}}=$ 60 MPa) is characterized by seismic events of complete ruptures, which propagate asymmetrically from one side of the seismogenic zone to the opposite side. (\textbf{c}) Model M3 ($p_{\text{eff}}=$ 20 MPa) displays that the same fault segment can regularly slip in slow and fast ruptures, resulting in recurrence intervals that vary over time. }
\label{modelM2M3}
\end{figure}

To investigate the influence that pore-fluid pressure level has on the spontaneous activation of \textit{seismic} and \textit{aseismic} slip on the fault, we perform further numerical experiments in which we decrease (model M2) and increase the effective pressure ($p_{\text{eff}}$) (model M3) compared to the value used for the reference model (Fig. \ref{modelM2M3}). The two models illuminate the effect of pore-fluid pressure throughout the earthquake cycle, from the recurrence period to the nucleation phase of seismic events, as well as the relation between \textit{seismic} and \textit{aseismic} slip. In model M2 (Fig. \ref{modelM2M3}b), due to a lower pore-fluid pressure, the effective pressure on the fault is higher ($p_{\text{eff}}=$ 60 MPa), and consequently the critical nucleation size decreases. As a result, seismic events nucleate earlier than in the reference model (Fig. \ref{eq-cycles}d). Furthermore, since we assume a slightly asymmetric fault structure (Fig. \ref{model_setup}b), seismic ruptures nucleate on one side of the seismogenic fault and propagate unilaterally on the other side of the fault segment in an asymmetric fashion. On the other hand, model M3 (Fig. \ref{modelM2M3}c) assumes a higher pore-fluid pressure, which results in a lower effective pressure ($p_{\text{eff}}=$ 20 MPa). In this case, the average shear strength is significantly lower than in model M2 and plastic yielding occurs more easily. Consequently, the seismogenic zone promotes unstable slip, either as seismic partial ruptures or as slow-slip transients. The long-term slip history indicates a mixture of fast and slow events, with a recurrence interval that vary over time. These results are important, as they suggest that pore-fluid pressure evolution in the same fault segment can release tectonic stress through both slow and fast ruptures, without incorporating any rate- and state-dependent friction.

\section{Discussion}\label{discussion}
\subsection{Poro-visco-elasto-plastic (de)compaction and propagation of solitary pressure waves}
A large body of theoretical work has explored the physics of solid-fluid interaction in active faults and earthquake source processes \citep[e.g.,][and references therein]{rice1976stabilization,rice1992fault,segall1995dilatancy,garagash2003shear,dunham2008earthquake,liu2010role,miller2013role}. In particular, a number of studies have focused on the relationship between porosity evolution, (de)compaction of reologically complex visco-(elasto)-plastic porous media, and fluid transport through the propagation of porosity waves \citep[e.g.,][]{connolly2000temperature,connolly2007decompaction,cruz2018rapid,yarushina2015compaction,skarbek2016dehydration,cruz2018rapid,petrini2020seismo}. \\

According to our results, it appears clear that visco-plastic compaction of fluid-filled porous media can lead to an abrupt self-pressurization of the fault zone, particularly under \textit{undrained} conditions or when the permeability on the fault is sufficiently low to prevent fluid flux on the boundaries of the fault zone on the timescale of seismic ruptures. As a result, the self-pressurization of the fault zone can trigger the propagation of solitary pressure waves, which can travel at co-seismic speed. To analyze the effect of solid-fluid (de)compaction, we quantify the separate contributions of visco-plastic compaction of the solid skeleton ($\zeta_{[\text{vp}]}$) and the decompaction of the fluid phase ($\zeta_{[\text{e}]}$) as follow: 

\begin{fleqn}[\parindent]
\begin{align} 
\zeta_{[\text{vp}]} = \frac{p^{[\text{t}]} - p^{[\text{f\,}]}}{\eta^{[\phi]} \, (1-\phi)}  \; ,
\label{comp-vp}
\end{align} 
\end{fleqn}

\begin{fleqn}[\parindent]
\begin{align} 
\zeta_{[\text{e}]} = \frac{\beta^{[\text{d}]} \, ((p_{| t}^{[\text{t}]} - p_{| t-\Delta_t}^{[\text{t}]})- \alpha (p_{| t}^{[\text{f\,}]} + p_{| t-\Delta_t}^{[\text{f\,}]}))}{\Delta_t}  \; .
\label{decomp-el}
\end{align} 
\end{fleqn}

\begin{figure}[h!]
\makebox[\textwidth][c]{\includegraphics[width=1.0\textwidth]{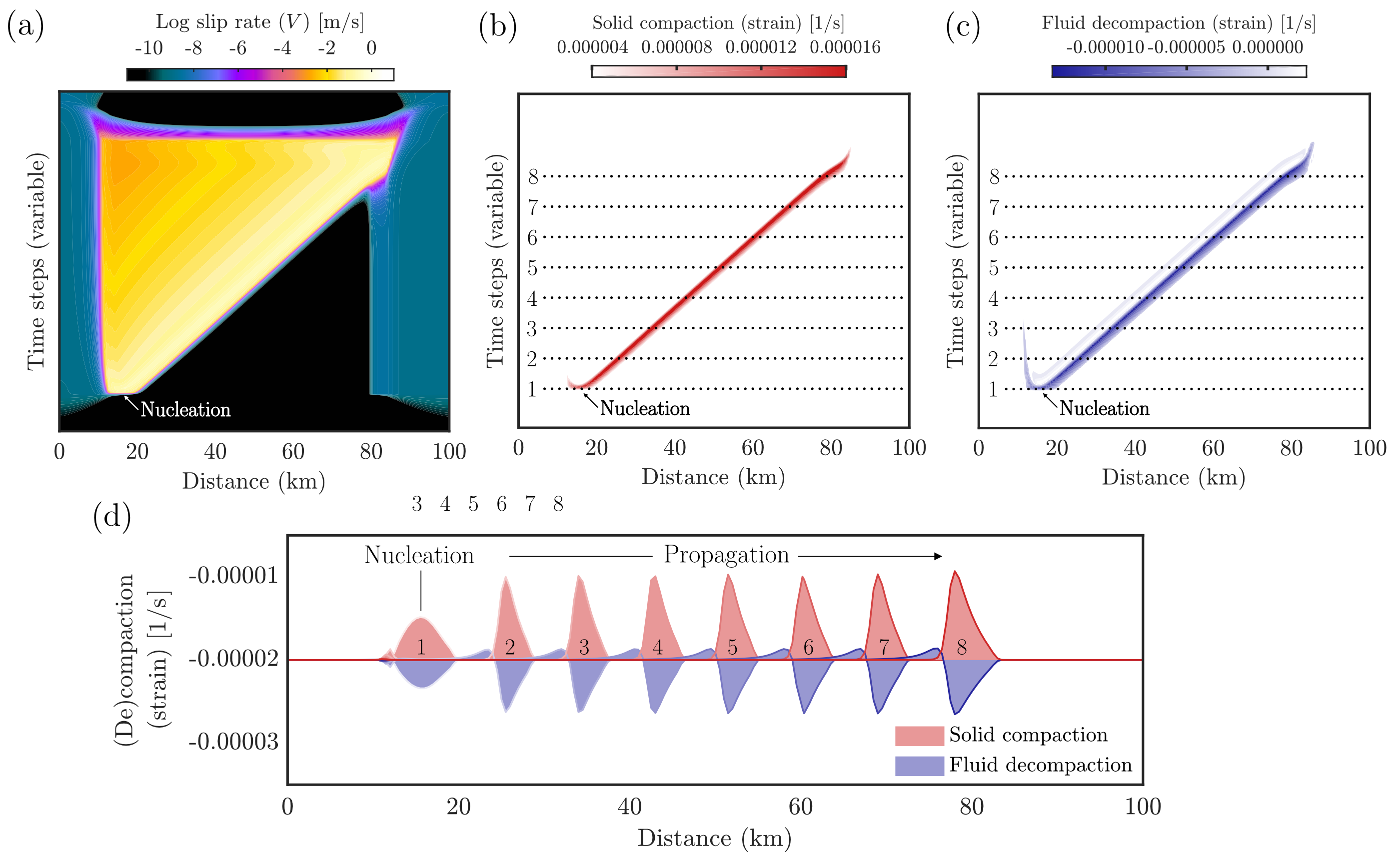}}
\caption{Compaction vs. decompaction within the fault zone. (\textbf{a}) Temporal evolution of slip rate for a representative major rupture events. (\textbf{b}) Evolution of visco-plastic compaction of the solid skeleton, computed from Eq. (\ref{comp-vp}). (\textbf{c}) Evolution of elastic decompaction of the fluid phase, computed from Eq. (\ref{decomp-el}). (\textbf{d}) Pressure wave propagation due to dynamic rupture, in which the visco-plastic compaction is mirrored by the elastic decompaction. The strain difference between visco-plastic compaction and elastic decompaction are computed from the 8 profiles shown in panel (\textbf{b}) and (\textbf{c}) and they represent the finite pore volume change.}
\label{decomp}
\end{figure}

(De)compaction along the fault during a representative major event is shown in Figure \ref{decomp}. Temporal evolution of the slip velocity displays a large event propagating unilaterally from the left to the right side of the seismogenic zone (Fig. \ref{decomp}a). An accurate quantification of visco-plastic compaction from Eq. (\ref{comp-vp}) and elastic decompaction from Eq. (\ref{decomp-el}) indicates that plastic yielding during the gradual nucleation phase is associated with a rapid increase in solid compaction (Fig. \ref{decomp}b), which is followed by a negative increase of fluid decompaction (Fig. \ref{decomp}c). As a result, pore-fluid pressure on the fault increases while shear strength decreases, thus causing a shear instability. Such shear instability grows on a relatively short time scale, causing the propagation of a dynamic rupture in the form of a pulse-like pressure wave (Fig. \ref{decomp}d). Notably, the strain difference between visco-plastic compaction and elastic decompaction represents the finite pore volume change. The combination of these two interconnected processes results in a small change in porosity and, in turn, permeability (Eq. \ref{perm-law}). Furthermore, a dynamic rupture causes an increased in pore-fluid pressure due to elastically stressed pores, which is better resolved by our fully compressible model with rate dependent strength compared to the previous simplified incompressible formulation \citep{petrini2020seismo}. Such increases in pore-fluid pressure are released over a relatively long diffusion period, which essentially depends on the permeability of both the host rock and the fault zone.  \\

\begin{figure}[h!]
\makebox[\textwidth][c]{\includegraphics[width=1.0\textwidth]{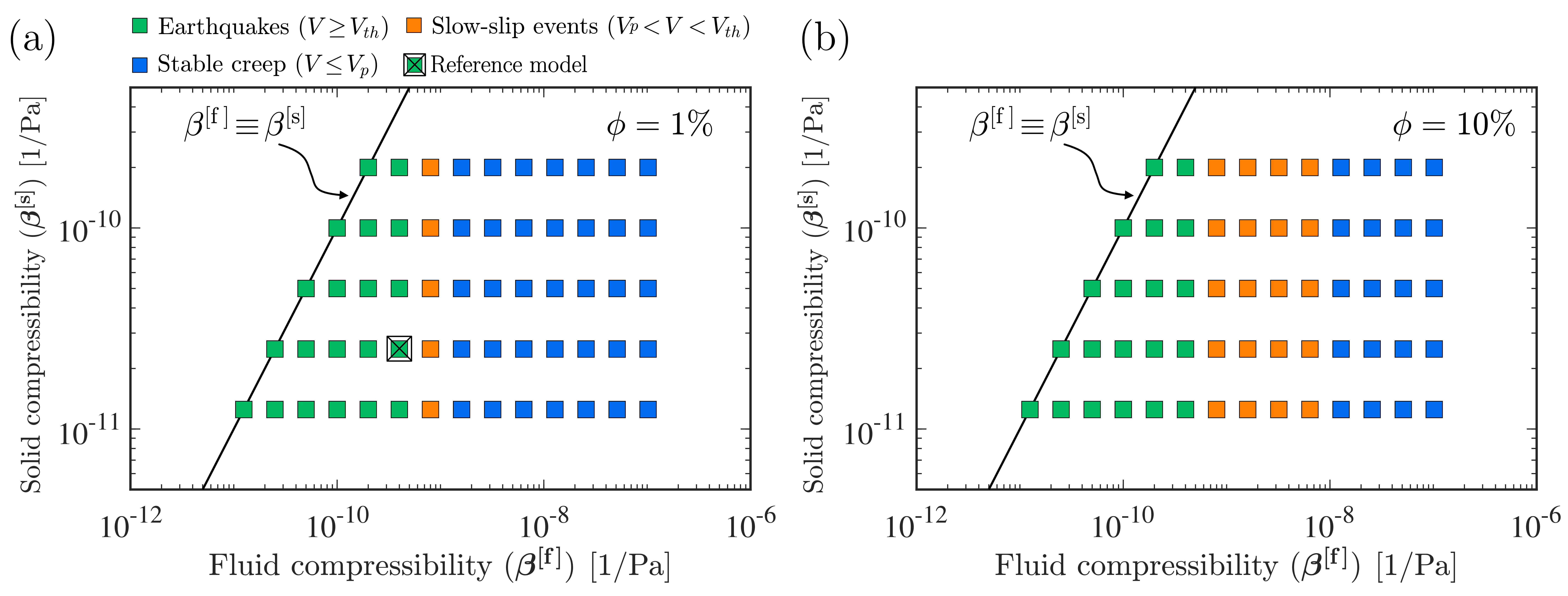}}
\caption{Impact of solid-fluid compressibility on the slip pattern. (\textbf{a}) Comparison between fluid compressibility ($\beta^{[\text{f\,}]}$) and solid compressibility ($\beta^{[\text{s}]}$) assuming a porosity ($\phi$) of 1$\%$. (\textbf{b}) Comparison between fluid compressibility and solid compressibility assuming a porosity of 10$\%$. Marker colors indicate the slip response of the fault, ranging from fast regular earthquake to slow-slip events and stable creep.}
\label{sol-flu-beta}
\end{figure}

It is important to note that the magnitude of visco-plastic and elastic (de)compaction depend on the compressibility of both the solid skeleton ($\beta^{[\text{s}]}$) and the fluid phase ($\beta^{[\text{f\,}]}$), as they have a direct impact on the Biot-Willis coefficient ($\alpha\,$; Eq. \ref{alpha}), and the Skempton coefficient ($B\,$; Eq. \ref{Skempton}). To quantify the role of solid and fluid compressibility, we execute two parameter studies of solid-fluid compressibility assuming two end-member porosities \citep[i.e., 1\% and 10\%; ][]{tewksbury2021constraints}, knowing that in nature the compressibility of porous rocks, and in particular fluids, can vary up to two orders of magnitude \citep[e.g.,][]{zimmerman1986compressibility,span1996new,mitchell2005fundamentals}. Our models produce distinctly different slip patterns within the range of solid-fluid compressibility (Fig. \ref{sol-flu-beta}). As illustrated in Figure \ref{sol-flu-beta}a, an increase in fluid compressibility leads to a transition from regular seismic events, to slow-slip events and eventually stable creep. In contrast, solid compressibility plays only a subordinate role. Furthermore, when we increase the reference porosity from 1\% to 10\% (Fig. \ref{sol-flu-beta}b), the parameter space in which we observe slow-slip events broaden. While high porosity serves as a damping factor to suppress the self-pressurization of pore-fluid and thus the occurrence of fast events, these results indicate that fluid compressibility can significantly affect the slip response of a fault loaded by tectonic stresses. As such, the higher the fluid compressibility, the lower the self-pressurization of the fault zone. Our numerical results can thus reproduce the full spectrum of fault slip behaviors under geophysically relevant conditions of pore-fluid pressure and fault composition, from slow to fast events, as observed for tectonic faults \citep[e.g.,][]{burgmann2018geophysics,obara2016connecting,jolivet2020transient}.

\subsection{Nucleation length, cohesive zone size, and fracture energy}
\begin{figure}[h!]
\makebox[\textwidth][c]{\includegraphics[width=1.0\textwidth]{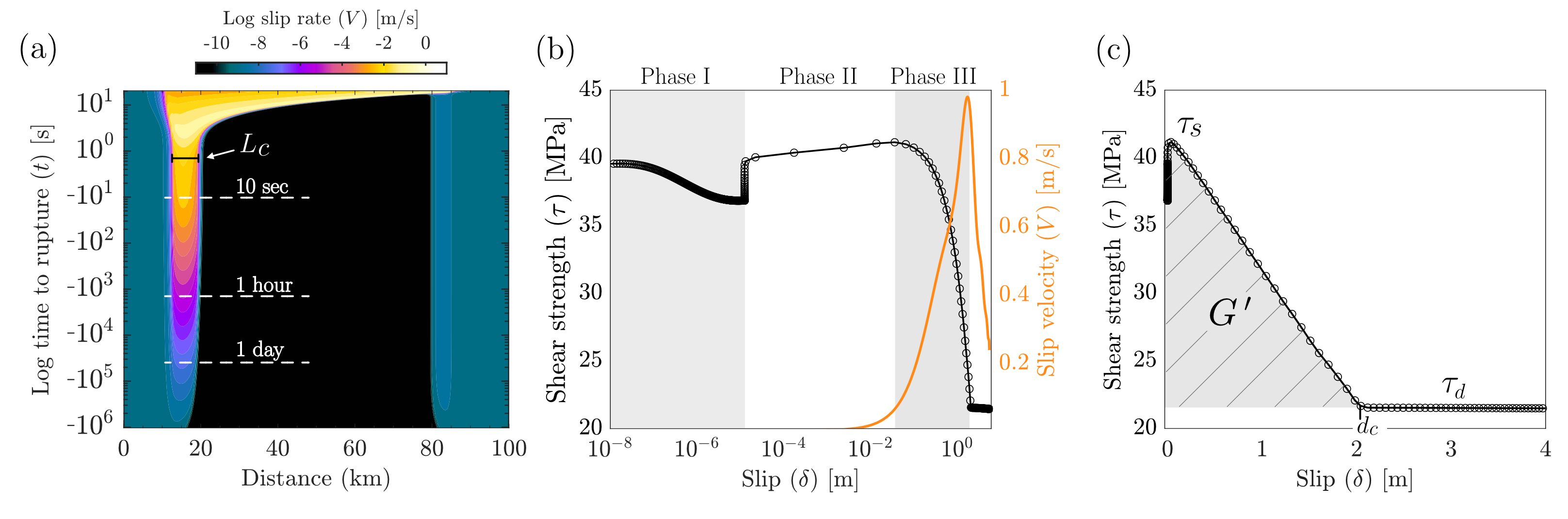}}
\caption{Nucleation size, slip evolution, and fracture energy. (\textbf{a}) Evolution of slip rate for a representative major event showing the gradual nucleation phase of a dynamic rupture. The $y$-axis displays time in log-scale. $L_c$ indicates the critical nucleation size (Eq. \ref{nucleation-size}).  (\textbf{b}) Weakening phases measured in the middle of the seismogenic zone ($x=50$ km). Black line and circles indicate the evolution of shear strength ($\tau$) with the cumulative slip ($\delta$), whereas the orange line displays the evolution of slip rate ($V$) with the cumulative slip. (\textbf{c}) Shear strength vs. slip. Shear strength linearly decreases from static shear strength ($\tau_s$) to a relatively low dynamic shear strength ($\tau_d$) over a characteristic slip weakening distance ($d_c$). The gray area illustrate the fracture energy ($G'$) computed from Eq. (\ref{Eq-fracture-energy}).}
\label{length-scale}
\end{figure}

The computational framework presented here is capable of resolving long-term deformation histories with continuous aseismic creep in the stable rate-strengthening fault regions throughout the loading period, nucleation of shear instabilities, co-seismic propagation of seismic ruptures, and post-seismic relaxation. As the considered example shown in Figure \ref{length-scale}, the algorithm is formulated with a rate-strengthening yield strength (Eq. \ref{syield}), in which the rate-strengthening exponent ($\gamma$) is decisive for its success to reproduce long-term inter-seismic periods with essentially quasi-static deformation, aseismic slip, and most importantly, gradual nucleation of dynamic ruptures (Fig. \ref{length-scale}a) \citep[e.g.,][]{dieterich1992earthquake,lapusta2000elastodynamic}. However, compared to the formulation of the classical rate- and state-dependent friction laws \citep[e.g.,][]{scholz1998earthquakes}, the evolutionary effect during the weakening phase is controlled by self-pressurization of pore-fluid pressure. \\

Our models indicate that shear cracks would become unstable when they reach the critical nucleation length ($L_c$) proposed by \citet{andrews1976rupture}, in which shear strength linearly decreases from static shear strength ($\tau_s$) to a relatively low dynamic shear strength ($\tau_d$) over a characteristic slip weakening distance ($d_c$) \citep{ida1972cohesive}

\begin{fleqn}[\parindent]
\begin{align} 
L_c = \frac{\mu \; (\tau_s-\tau_d) \; d_c}{\pi \; (1-\nu) \; (\tau_0-\tau_d)^2}\ \; ,
\label{nucleation-size}
\end{align} 
\end{fleqn}

where $\tau_0$ represents the initial shear stress (Eq. \ref{DruckerPrager}). According to our model (Fig. \ref{length-scale}a), Eq (\ref{nucleation-size}) predicts a critical nucleation length of $\sim$7.3 km. Comparing Fig. \ref{fig-slip-history}a and Fig. \ref{modelM2M3}a with Fig. \ref{length-scale}a, we notice that slip during and right after the nucleation phase are very similar despite the different critical nucleation length. This means that observing signals from the nucleation of fluid-driven shear cracks do not provide any evidence on the final size of the event. This suggests that the final size of seismic events is determined by the conditions on the fault, rather than by the nucleation process itself. \\

Figure \ref{length-scale}b illustrates the weakening phases that buffer the pore pressure and slip velocity rise, leading to dynamic rupture. During Phase I, upon rapid loading, shear stress rises abruptly with strain as the fault is elastically loaded previous to the onset of slip. Phase II is characterized by a strain-hardening shear strength, which produces an early phase of slip ($<$1 cm of slip), and corresponds roughly to slip rates of the order of 1--10 cm s$^{-1}$. During Phase III, an abrupt weakening is initiated and shear strength drops while slip rate quickly accelerates up to $\sim$m/s. Peak in slip rate is reached when strength drop is completed, which is follow by a quasi steady-state phase where low shear strength is maintained with minor fluctuations. \\

When an event nucleates, shear stress drops while slip velocity increases rapidly. Since the rupture speed at this stage is still close to zero, the length along which the stress drop occurs represents the quasi static cohesive zone length ($\Lambda$). The cohesive zone size is an important resolution criterion in dynamic rupture because it gives the spatial length scale over which the shear stress drops from its peak to residual values at the propagating rupture front \citep{palmer1973growth,day2005comparison}. Since fluid-driven shear cracks essentially mimic a linear slip-weakening law (Fig. \ref{length-scale}c), the cohesive zone sizes at rupture speed $c=0^+$  ($\Lambda_0$) observed in our simulations correspond quite well to the estimate proposed by \citet{day2005comparison}:

\begin{fleqn}[\parindent]
\begin{align} 
\Lambda_0 = \: \frac{9 \pi}{32} \: \frac{\mu^* \; d_c}{(\tau_s-\tau_d)} \: ,
\label{Eq_coh_zone}
\end{align} 
\end{fleqn}

where $9 \pi/32$ is a constant if the stress traction distribution within the cohesive zone is linear in space \citep{palmer1973growth}. This latter study establishes that $\Lambda_0$/$\Delta x$ of 3 to 5 is required to resolve dynamic rupture. The ratio of nucleation zone size and cohesive zone size ($\Lambda_0$/$L_c$) is between 0.3 and 0.5, as reported in previous studies \citep[e.g.,][]{lapusta2000elastodynamic}. Therefore, resolving the cohesive zone is the more stringent numerical criterion here. In our models, we choose the cell size small enough to resolve the cohesive zone of 2.5 km with at least 10 cells. Hence the spatial discretization in the simulations, with the cell size of $\Delta x$=250 m, is small enough to resolve the evolution of stress and slip rate. \\

Seismic ruptures are controlled by an energy balance involving elastic work, wave radiation, and dissipation by anelastic processes, including friction and plastic strain. As illustrated in Figure \ref{length-scale}c, the frictional work during the weakening process --- as the product of shear stress and slip --- equates to a simple form of fracture energy ($G'$), which assumes a linear slip weakening \citep{palmer1973growth,andrews1976rupture,okubo1984effects}

\begin{fleqn}[\parindent]
\begin{align} 
G' (\delta) = \int_{0}^{\delta} [ \, \tau (\delta') - \tau (\delta) \, ] \, \text{d}\delta'  \; \approx \; \frac{1}{2} \; (\tau_s-\tau_d) \; d_c \, .
\label{Eq-fracture-energy}
\end{align} 
\end{fleqn}

\begin{figure}[h!]
\makebox[\textwidth][c]{\includegraphics[width=1.0\textwidth]{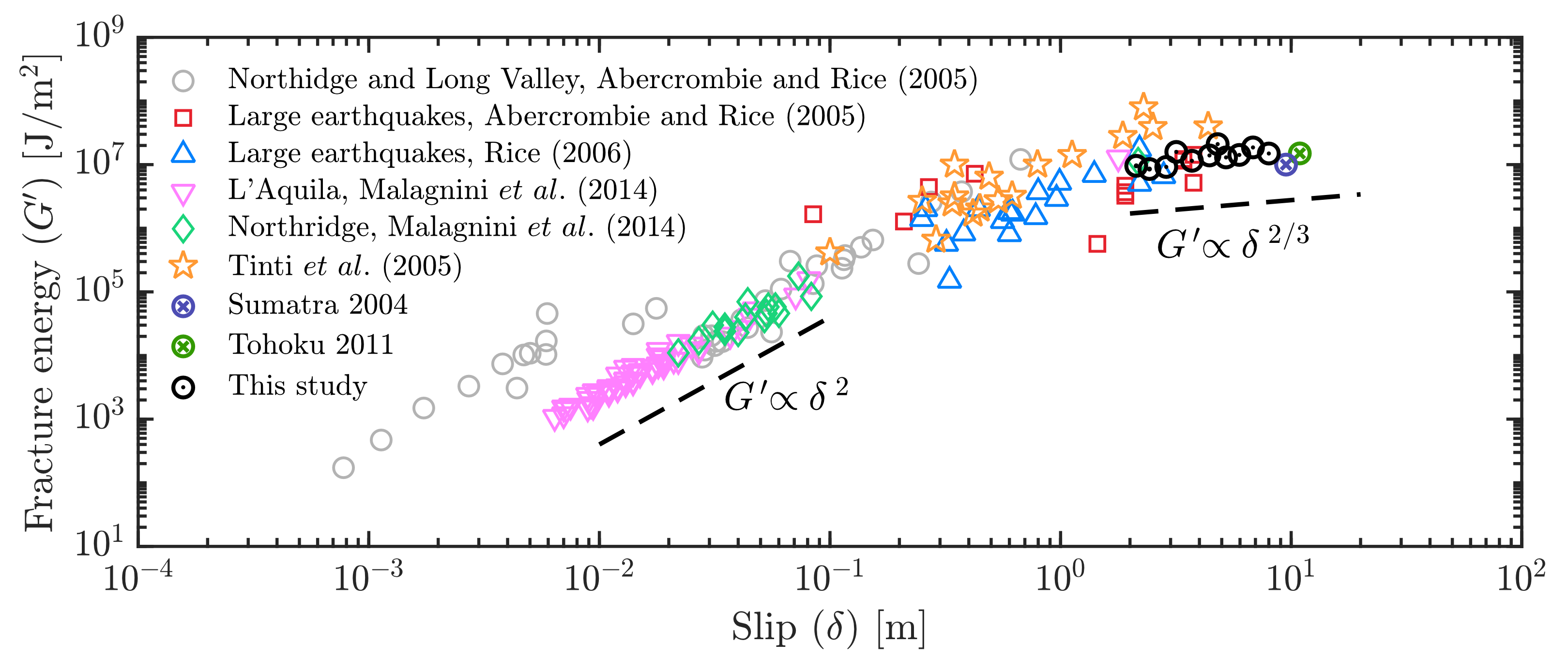}}
\caption{Compilation of fracture energy ($G'$) and slip ($\delta$) from different earthquakes worldwide, from borehole microseismicity to great earthquakes \citep{abercrombie2005can,rice2006heating,malagnini2014gradual,viesca2015ubiquitous,nielsen2016scaling,tinti2005earthquake}. Black dashed lines indicate the different scaling for small ($G\propto \delta{\,^2}$) and large ($G\propto \delta{\,^{2/3}}$) earthquakes.}
\label{fracture-energy}
\end{figure}

To compare our modeling results with existing measurements of fracture energy from earthquakes globally, we compile fracture energy estimates made previously with events spanning five orders of magnitude in size, from borehole microseismicity to great earthquakes \citep{abercrombie2005can,rice2006heating,malagnini2014gradual,viesca2015ubiquitous,nielsen2016scaling,tinti2005earthquake} (Fig. \ref{fracture-energy}). Previous inferences have shown a nonlinear scaling of fracture energy with slip \citep{abercrombie2005can}, from $G\propto \delta{\,^2}$ for small earthquakes to $G\propto \delta{\,^{2/3}}$ for large earthquakes (Fig. \ref{fracture-energy}), which has been attributed to thermal pressurization of pore-fluid by the rapid shear heating of fault gouge \citep{viesca2015ubiquitous}. For the set of parameters considered in this study, our models produce both relatively large stress drop ($\tau_s-\tau_d$) and large characteristic slip weakening distance ($d_c$), which results in high values of fracture energy (Fig. \ref{length-scale}c). Remarkably, the range of fracture energy and slip from our dynamic analysis supports the observed $G\propto \delta{\,^{2/3}}$ scaling of fracture energy for large earthquakes, and suggests that self-pressurization of fluids is a viable mechanism for explaining widespread and prominent process of fault weakening. \\

It is important to note that in our models $d_c$ is not a constant, as it is treat as variable and it is dynamically determined during rupture itself (Fig. \ref{length-scale}c). The weakening distance $d_c$ has been widely used in fault studies and often imposed as a constant \citep[e.g.,][]{tse1986crustal,ben1995slip,lapusta2000elastodynamic,rubin2005earthquake,lapusta2009three}. However, ample observations suggest that $d_c$ should be treated as a variable \citep{cocco2002slip,nielsen2010transient}, since the effective weakening distance depends on slip history and loading conditions \citep[e.g.,][]{guatteri2001inferring,tinti2004estimates}. Furthermore, laboratory measures of $d_c$ using rotary friction experiments performed at seismic slip rates ($\sim1$ m s$^{-1}$) yield weakening distances of the order of meters \citep[][and references therein]{nielsen2016scaling}, substantial friction drops \citep{di2011fault}, and fracture energies arguably in the same order as the seismological estimates \citep{nielsen2016g}.

\subsection{Future avenues for research}
Future studies will make use of this methodology to investigate a number of different phenomena, either as a natural process or induced by human activities. The proposed setup is rather simple, but it can be applied to investigate, e.g., fluid-driven swarms and foreshock sequences \citep[e.g.,][]{ross20203d,miller2020aftershocks}. The developed methodology presents a promising avenue for investigating slip that occurs in response to fluid injection into a permeable fault governed by either rate-strengthening yield strength or rate-and-state friction \citep[e.g.,][]{yang2021effect}. Future work will also focus on more realistic scenarios of geologic faults, which will allow to investigate the full spectrum of slip behaviour observed in subduction zones and strike-slip faults \citep[e.g.,][]{jolivet2020transient,behr2021s}. \\

Over the last decades, seismic and geodetic observations have revealed complex interactions of earthquakes, slow-slip transients, and stable creep \citep{burgmann2018geophysics}. Investigating whether pore-fluid diffusion affects variations in earthquake recurrence intervals, or fault tremor sequences with alternating longer and shorter repeat times (so-called period-doubling events) \citep{shelly2010periodic}, represents an important topic in earthquake physics. Also, systematic investigation of the conditions under which slow-slip events are mostly controlled by visco-plastic bulk properties \citep[e.g.,][]{viesca2018slow,lambert2016contribution,gao2017rheological,fagereng2021complex,velo2cycles}, rather than fault frictional properties \citep[e.g.,][]{liu2007spontaneous,dal2020unraveling}, is an important frontier in earthquake source modeling. Further developments can also incorporate fault dilation, temperature, shear heating, thermal runaway, and grain size evolution \citep[e.g.,][]{austin2007paleowattmeters}, which is needed to investigate the onset of highly localized viscous creep in pre-existing, fine-grained shear zones. As temperature may vary markedly with shear heating \citep{viesca2015ubiquitous}, the temperature-weakening behavior of steady-state friction may offer the conditions for episodic slow-slip transients, even under rate-strengthening fault behavior.

\section{Conclusions}

We have developed a finite difference method to account for fully coupled solid-fluid interaction over many earthquake cycles. The computational framework can model rate-strengthening plasticity in a poro-visco-elasto-plastic rheology. Numerical results are verified through convergence tests and comparisons with analytical benchmarks of pore-fluid pressure diffusion from an injection point along a finite fault width. Future work will include a deeper exploration of parameter space, including the effects of permeability, rate-strengthening, and viscosity, as our set of parameters were chosen primarily for efficiency of computation. For the parameter study in this work, we found that pore-fluid pressure, porosity, and solid-fluid compressibility, influence the occurrence of \textit{seismic} and \textit{aseismic} slip. \\

We have investigated spontaneously occurring shear instabilities on a mildly rate-strengthening fault zone with solid-fluid coupling. Remarkably, these instabilities are fundamentally different from standard instabilities with rate-and-state friction in that they are controlled by localized (de)compaction of pores and dynamic self-pressurization of fluids inside the undrained fault zone. Simulations show that these fluid-driven dynamic ruptures are controlled by solitary pulse-like pressure waves propagating at seismic speed. We have proposed a conceptual model for how this type of instability manifest in a 2-D plane-strain shear model (mode-II). However, we recognize that in order to make a full comparison to geological settings we need to understand how the reported instabilities manifest in three dimensions. Despite this limitation, our work demonstrates how pore-fluid pressure, poroelastic effects, and inelastic deformation, can destabilize active faults and produce the full spectrum of slip, from fast- to slow-slip, consistent with observations. \\

Although multiple weakening mechanisms may operate on active faults, our results suggest that self-pressurization of fluid-filled porous rocks may be the dominant contributor to fault fracture energy, particularly for large earthquakes. Understanding how faults lose their shear strength during fluid-induced dynamic rupture is critical, as it may help to constrain the minimum level of shear stress a fault requires to become unstable. In a broader context, this study shows the importance of incorporating the realistic hydro-mechanical structure of faults to investigate sequences of \textit{seismic} and \textit{aseismic} slip. In particular, this study indicates that pore-pressure evolution can completely change both the failure process on the interface and the long-term slip history of geologic faults.


\section*{Declaration of Competing Interest}
None.

\section*{Acknowledgments}
We thank Nadia Lapusta, Massimo Cocco, Paul Selvadurai, Elias Heimisson, Federico Ciardo, Domenico Giardini, and Casper Pranger for discussions. This research was supported by the European Research Council (ERC) Synergy Grant FEAR (ID: \texttt{856559}). Numerical simulations were performed on ETH cluster Euler.  \\

\section*{Data availability}
\noindent The data underlying this article will be shared upon request to the corresponding author.\\

\section*{Code availability}
\noindent Computer code used within the manuscript is still under development and is available from the corresponding author upon request.

\appendix{
\section{Discretization of the governing equations}\label{discretization}
\noindent The governing equations (\ref{mom-sol+flu}--\ref{mass-flu}) are solved on a 2-D fully staggered Eulerian grid with $N_x$ and $N_y$ number of nodes in the $x-$ and $y-$direction respectively. The different variables are fully staggered on the grid as shown in Fig. \ref{fig-grid} to ensure optimal discretization. Figure \ref{grid_eq_1} shows the stencils used to discretize in 2-D the total momentum (solid matrix and fluid; Eq. \ref{mom-sol+flu}) under plane strain conditions ($v_z=0$) for $x-$ and $y-$components by including Eqs. (\ref{inf-strain}) and (\ref{full-stress}):

\begin{fleqn}[\parindent]
\begin{align}
\begin{split}
& \, \bigg(\frac{4}{3}\frac{\eta^*_{[xx](i,\, j+1)} \, (v^{[\text{s}]}_{x(i,\, j+1)} -v^{[\text{s}]}_{x(i,\, j)})}{\Delta x^2} - \; \frac{4}{3} \frac{\eta^*_{[xx](i,\, j)} \, (v^{[\text{s}]}_{x(i,\, j)} -v^{[\text{s}]}_{x(i,\, j-1)})}{\Delta x^2} \bigg) \; .\,.\,.\\ 
- &  \bigg(\frac{2}{3} \frac{\eta^*_{[xx](i,\, j+1)} \, (v^{[\text{s}]}_{y(i,\, j+1)} -v^{[\text{s}]}_{y(i-1,\, j+1)})}{\Delta x \; \Delta y} -\; \frac{2}{3} \frac{\eta^*_{[xx](i,\, j)} \, (v^{[\text{s}]}_{y(i,\, j)} -v^{[\text{s}]}_{y(i-1,\, j)})}{\Delta x \; \Delta y} \bigg) \; .\,.\,.\\
+& \, \bigg(\frac{\eta^*_{[xy](i,\, j)} \, (v^{[\text{s}]}_{x(i+1,\, j)} -v^{[\text{s}]}_{x(i,\, j)})}{\Delta y^2} - \frac{\eta^*_{[xy](i-1,\, j)} \, (v^{[\text{s}]}_{x(i,\, j)} -v^{[\text{s}]}_{x(i-1,\, j)})}{\Delta y^2} \bigg) \; .\,.\,.\\
+ & \, \bigg(\frac{\eta^*_{[xy](i,\, j)} \, (v^{[\text{s}]}_{y(i,\, j+1)} -v^{[\text{s}]}_{y(i,\, j)})}{\Delta y \; \Delta x} - \frac{\eta^*_{[xy](i-1,\, j)} \, (v^{[\text{s}]}_{y(i-1,\, j+1)} -v^{[\text{s}]}_{y(i-1,\, j)})}{\Delta y \; \Delta x} \bigg) - \frac{p_{(i,\, j+1)}^{[\text{t}]} - p_{(i,\, j)}^{[\text{t\,}]}}{\Delta x} + \rho^{[\text{t}]} \, \boldsymbol{g}_{x} \\
= & \; \rho^{[\text{t}]} \; \frac{v_{x(i,\, j)\, | \, t}^{[\text{s}]} - v_{x(i,\, j) \, | \, t-\Delta_t }^{[\text{s}]}}{\Delta t} - \bigg( \frac{\tau_{xx \,(i,\, j+1)}^* - \tau_{xx \,(i,\, j)}^*}{\Delta x} \bigg) - \bigg( \frac{\tau_{xy \,(i,\, j)}^* - \tau_{xy \,(i-1,\, j)}^*}{\Delta y} \bigg)
\end{split}
\end{align}
\end{fleqn}

\begin{fleqn}[\parindent]
\begin{align}
\begin{split}
& \bigg(\frac{4}{3} \frac{\eta^*_{[yy](i+1,\, j)} \, (v^{[\text{s}]}_{y(i+1,\, j)} -v^{[\text{s}]}_{y(i,\, j)})}{\Delta y^2} - \; \frac{4}{3} \frac{\eta^*_{[yy](i,\, j)} \, (v^{[\text{s}]}_{y(i,\, j)} -v^{[\text{s}]}_{y(i-1,\, j)})}{\Delta y^2} \bigg) \; .\,.\,.\\ 
-&  \bigg(\frac{2}{3} \; \frac{\eta^*_{[yy](i+1,\, j)} \, (v^{[\text{s}]}_{x(i+1,\, j)} -v^{[\text{s}]}_{x(i+1,\, j-1)})}{\Delta y \; \Delta x} - \frac{2}{3}\; \frac{\eta^*_{[yy](i,\, j)} \, (v^{[\text{s}]}_{x(i,\, j)} -v^{[\text{s}]}_{x(i,\, j-1)})}{\Delta y \; \Delta x} \bigg) \; .\,.\,.\\
+ & \bigg(\frac{\eta^*_{[xy](i,\, j)} \, (v^{[\text{s}]}_{x(i+1,\, j)} -v^{[\text{s}]}_{x(i,\, j)})}{\Delta x \; \Delta y} - \frac{\eta^*_{[xy](i,\, j-1)} \, (v^{[\text{s}]}_{x(i+1,\, j-1)} -v^{[\text{s}]}_{x(i,\, j-1)})}{\Delta x \; \Delta y} \bigg) \; .\,.\,.\\
+ & \bigg(\frac{\eta^*_{[xy](i,\, j)} \, (v^{[\text{s}]}_{y(i,\, j+1)} -v^{[\text{s}]}_{y(i,\, j)})}{\Delta x^2} - \frac{\eta^*_{[xy](i,\, j-1)} \, (v^{[\text{s}]}_{y(i,\, j)} -v^{[\text{s}]}_{y(i,\, j-1)})}{\Delta x^2} \bigg) - \frac{p_{(i+1,\, j)}^{[\text{t}]} - p_{(i,\, j)}^{[\text{t\,}]}}{\Delta y} + \rho^{[\text{t}]} \, \boldsymbol{g}_{y} \\
= & \; \rho^{[\text{t}]} \; \frac{v_{y(i,\, j)\, | \, t}^{[\text{s}]} - v_{y(i,\, j) \, | \, t-\Delta_t }^{[\text{s}]}}{\Delta t} - \bigg( \frac{\tau_{yy \,(i,\, j+1)}^* - \tau_{yy \,(i,\, j)}^*}{\Delta y} \bigg) - \bigg( \frac{\tau_{xy \,(i,\, j)}^* - \tau_{xy \,(i,\, j-1)}^*}{\Delta x} \bigg)
\end{split}
\end{align}
\end{fleqn}

where $\eta^*= \eta_{\text{vp}} \; Z$, $\tau^*= \tau \; (1- Z)$. \\

Figure \ref{grid_eq_24}a,b displays the stencils used to discretize the Darcy Eq. (\ref{mom-flu}), whereas Figure \ref{grid_eq_24}c,d shows the stencils used to discretize the fully compressible solid mass (Eq. \ref{mass-sol}) and and the fully compressible fluid mass (Eq. \ref{mass-flu-stencil}), respectively:

\begin{fleqn}[\parindent]
\begin{align}
\boldsymbol{v}_x{(i,j)}^{[\text{D}]} = - \frac{k_{(i,\, j)}^{[\phi]}}{\eta_{(i,\, j)}^{[\text{f\,}]}} \; \bigg(\frac{p_{(i,\, j+1)}^{[\text{f\,}]} - p_{(i,\, j)}^{[\text{f\,}]}}{\Delta x} - \rho^{[\text{f\,}]} \, \bigg( \boldsymbol{g}_x - \; \frac{v_{x(i,\, j)\, | \, t}^{[\text{f\,}]} - v_{x(i,\, j) \, | \, t-\Delta_t }^{[\text{f\,}]}}{\Delta t} \bigg)\bigg)
\label{mom-flu-stencil-a} 
\end{align}
\end{fleqn}

\begin{fleqn}[\parindent]
\begin{align}
\boldsymbol{v}_y{(i,j)}^{[\text{D}]} = - \frac{k_{(i,\, j)}^{[\phi]}}{\eta_{(i,\, j)}^{[\text{f\,}]}} \; \bigg(\frac{p_{(i+1,\, j)}^{[\text{f\,}]} - p_{(i,\, j)}^{[\text{f\,}]}}{\Delta y} - \rho^{[\text{f\,}]} \, \bigg( \boldsymbol{g}_y - \; \frac{v_{y(i,\, j)\, | \, t}^{[\text{f\,}]} - v_{y(i,\, j) \, | \, t-\Delta_t }^{[\text{f\,}]}}{\Delta t} \bigg)\bigg)
\label{mom-flu-stencil-b}  
\end{align}
\end{fleqn}

\begin{fleqn}[\parindent]
\begin{align}
\begin{split}
&  \, \bigg(\frac{(v^{[\text{s}]}_{x(i,\, j)} -v^{[\text{s}]}_{x(i,\, j-1)})}{\Delta x} + \frac{(v^{[\text{s}]}_{y(i,\, j)} -v^{[\text{s}]}_{y(i-1,\, j)})}{\Delta y}  \bigg) \; .\,.\,. \\ 
&  = - \frac{1}{K_{(i,\, j)}^{[\text{d}]}} \; \bigg(\frac{p_{(i,\, j)\, | \, t}^{[\text{t}]} - p_{(i,\, j) \, | \, t-\Delta_t }^{[\text{t}]}}{\Delta t} - \alpha_{(i,\, j)} \frac{p_{(i,\, j)\, | \, t}^{[\text{f\,}]} - p_{(i,\, j) \, | \, t-\Delta_t }^{[\text{f\,}]}}{\Delta t}\bigg) - \frac{p_{(i,\, j)}^{[\text{t}]} - p_{(i,\, j)}^{[\text{f\,}]}}{\eta_{(i,\, j)}^{[\phi]}(1-\phi_{(i,\, j)})} \, 
\label{mass-sol-stencil} 
\end{split}
\end{align}
\end{fleqn}

\begin{fleqn}[\parindent]
\begin{align}
\begin{split}
&  \, \bigg(\frac{(v^{[\text{D}]}_{x(i,\, j)} -v^{[\text{D}]}_{x(i,\, j-1)})}{\Delta x} + \frac{(v^{[\text{D}]}_{y(i,\, j)} -v^{[\text{D}]}_{y(i-1,\, j)})}{\Delta y}  \bigg) \; .\,.\,.\\ 
&  = \frac{\alpha_{(i,\, j)}}{K_{(i,\, j)}^{[\text{d}]}} \; \bigg(\frac{p_{(i,\, j)\, | \, t}^{[\text{t}]} - p_{(i,\, j) \, | \, t-\Delta_t }^{[\text{t}]}}{\Delta t} - \frac{1}{B_{(i,\, j)}} \frac{p_{(i,\, j)\, | \, t}^{[\text{f\,}]} - p_{(i,\, j) \, | \, t-\Delta_t }^{[\text{f\,}]}}{\Delta t}\bigg) + \frac{p_{(i,\, j)}^{[\text{t}]} - p_{(i,\, j)}^{[\text{f\,}]}}{\eta_{(i,\, j)}^{[\phi]}(1-\phi_{(i,\, j)})} \, 
\label{mass-flu-stencil} 
\end{split}
\end{align}
\end{fleqn}

\begin{figure}[h!]
\makebox[\textwidth][c]{\includegraphics[width=1.0\textwidth]{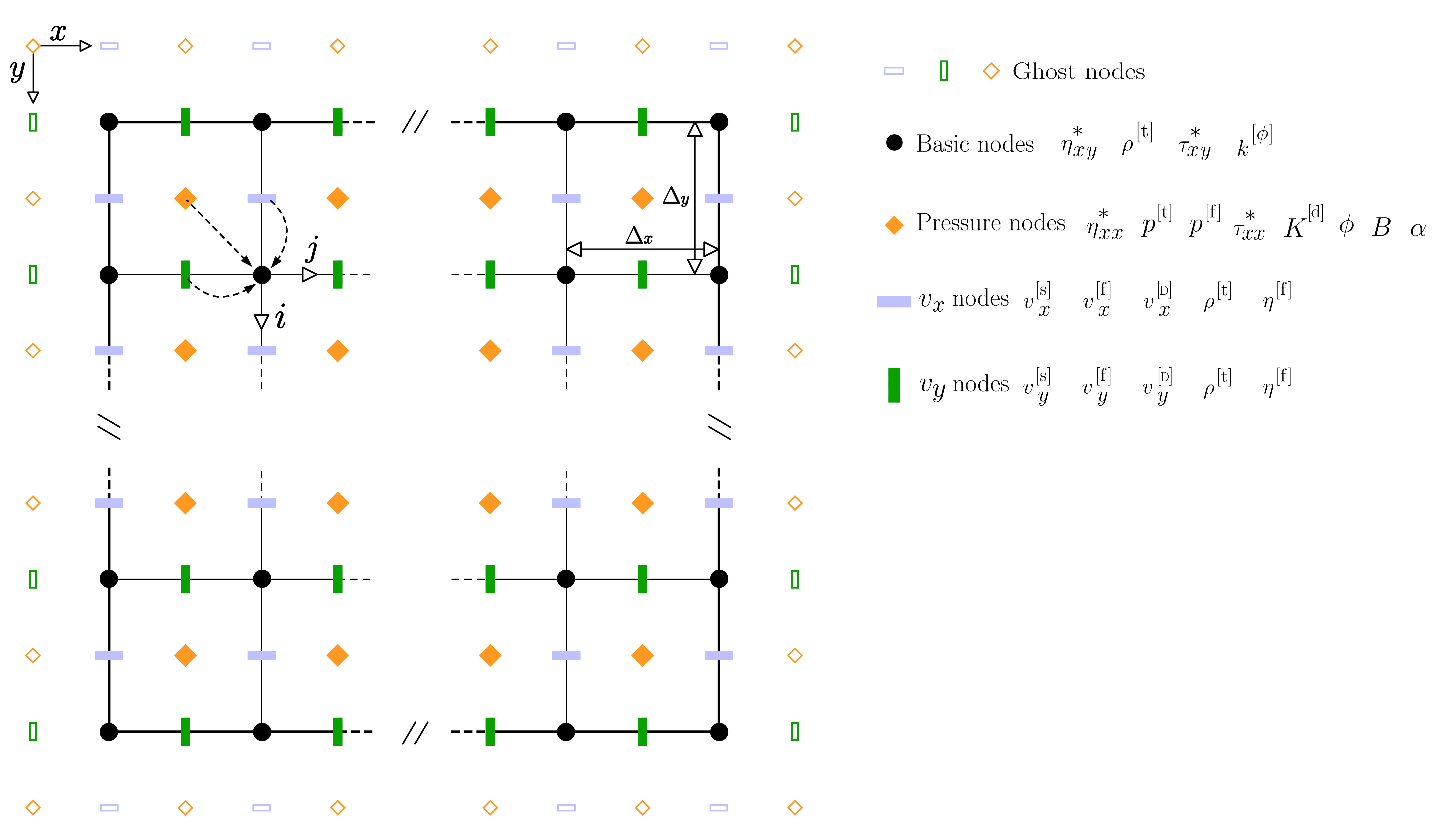}}
\caption{Fully staggered grid with representation of the different nodes. The ghost nodes are not used to solve the governing equations and boundary conditions. Different node symbols and colors illustrate the distribution of variables on the staggered nodes.}
\label{fig-grid}
\label{grid}
\end{figure}

\begin{figure}[h!]
\makebox[\textwidth][c]{\includegraphics[width=0.9\textwidth]{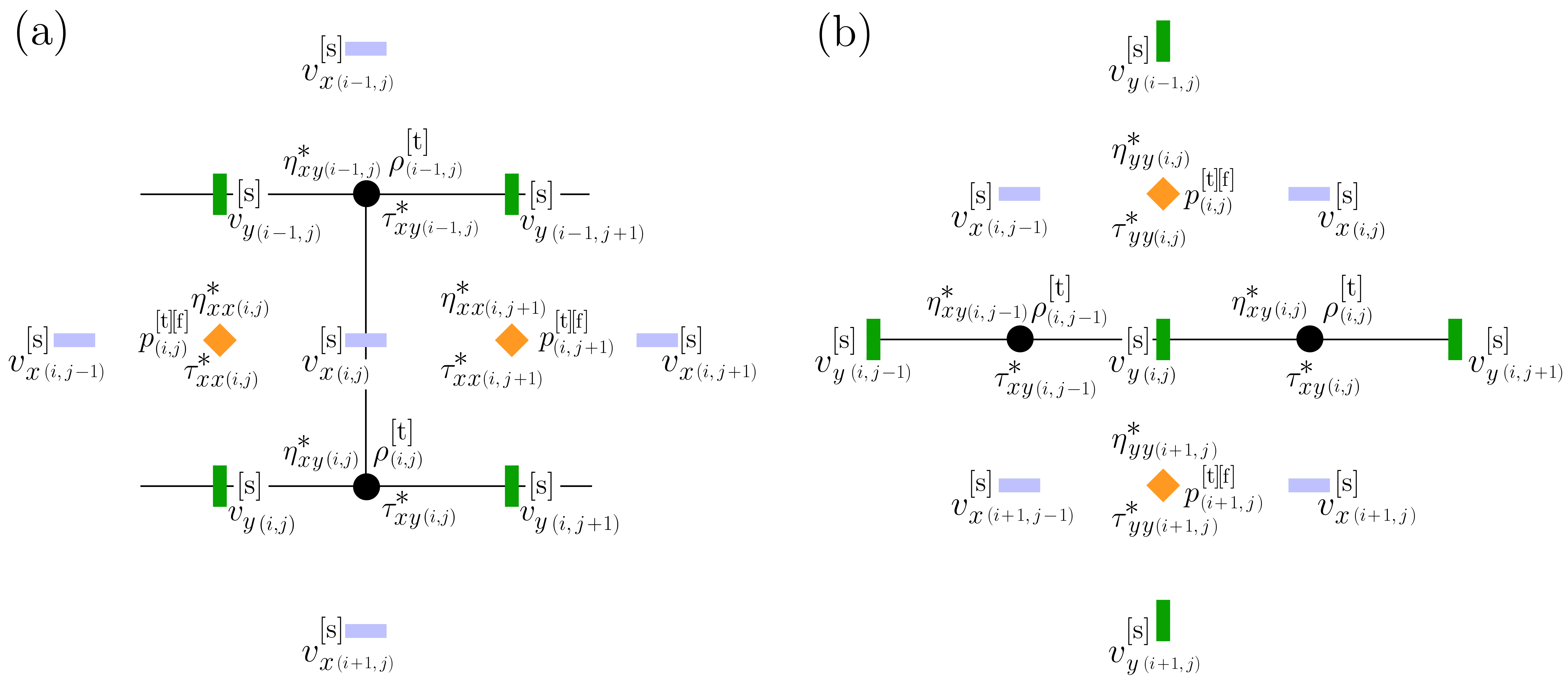}}
\caption{Staggered-grid stencils used by our scheme to discretize the total momentum (solid matrix and fluid; Eq. \ref{mom-sol+flu}) for $x-$ and $y-$components.}
\label{grid_eq_1}
\label{grid}
\end{figure}

\begin{figure}[h!]
\makebox[\textwidth][c]{\includegraphics[width=0.8\textwidth]{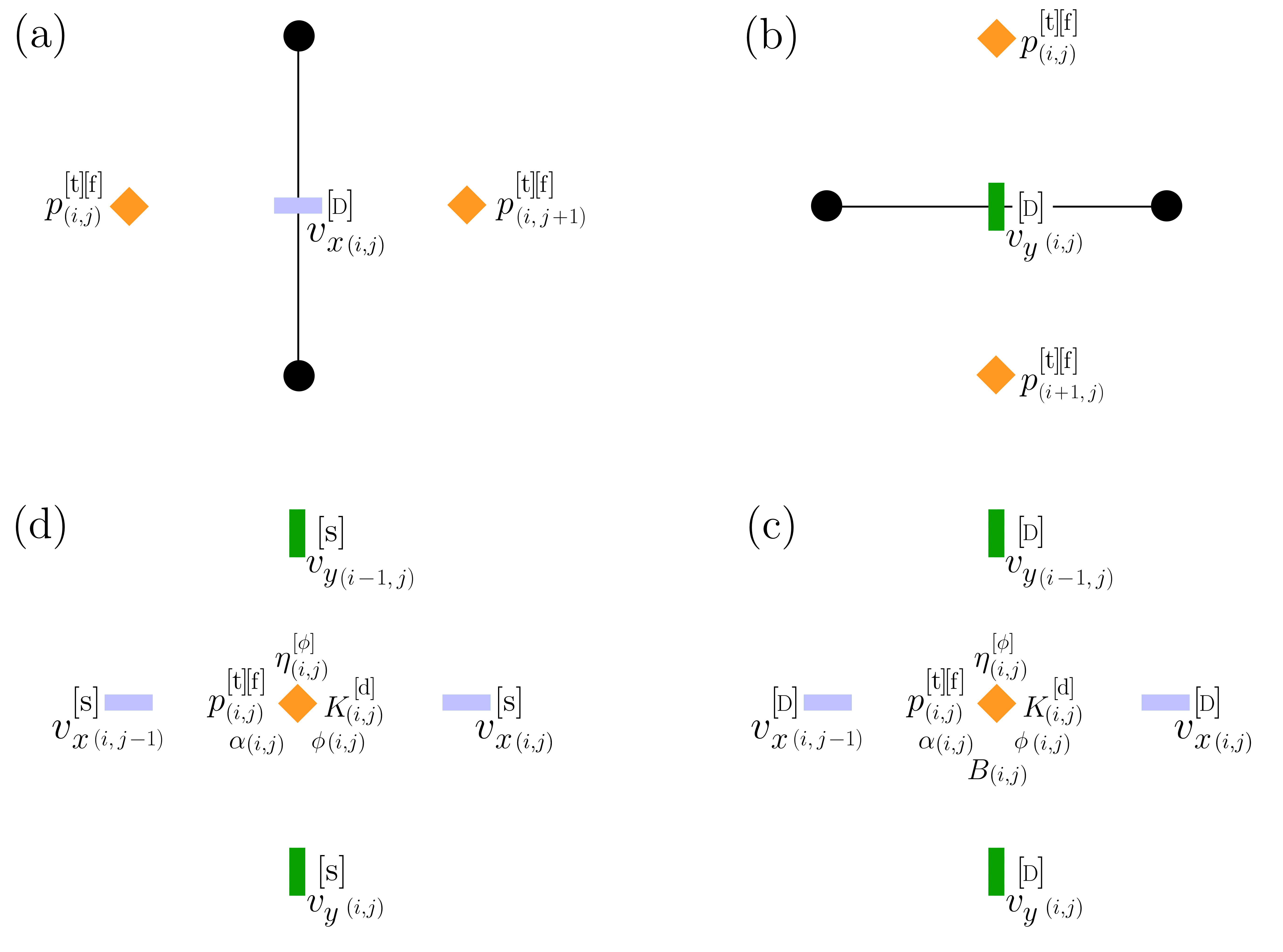}}
\caption{Stencils used for discretizing the Darcy equations: (\textbf{a}) $x$-Darcy equation and (\textbf{b}) $y$-Darcy equation (Eq. \ref{mom-flu}). Stencils used for discretizing the conservation equations: (\textbf{c}) solid mass (Eq. \ref{mass-sol}) and (\textbf{d}) fluid mass (Eq. \ref{mass-flu}).}
\label{grid_eq_24}
\label{grid}
\end{figure}
}

\clearpage

\newpage

\bibliographystyle{apalike}
\bibliography{../references/refs}

\end{document}